\documentclass[fleqn,10pt]{article}
\usepackage{latexsym, graphicx, epsfig, amsmath, notation, amssymb,amsfonts}
\usepackage{natbib,amsthm,version}
\usepackage{amsbsy,bm,multirow}
\usepackage[titletoc,page]{appendix}
\usepackage[mathscr]{eucal}
\usepackage{enumerate,color}

\let\emptyset\varnothing

\title{
A Convective-like Energy-Stable Open Boundary Condition
for Simulations of Incompressible Flows
} 
\author{
  S. Dong\thanks{Email: sdong@purdue.edu} \\
  Center for Computational \& Applied Mathematics \\
  Department of Mathematics \\
  Purdue University, USA
 } 

\date{} %(\today)}
\begin{document}
\maketitle

%% double space
\baselineskip 2em %2.2em

%%%%%%%%%%%%%%%%%%%%%%%%%%%%%%%%%%%%%%%%%%%%%%%%%%
%% Abstract

\begin{abstract}

We present a new energy-stable open boundary
condition, and an associated numerical algorithm, 
for simulating incompressible flows
with outflow/open boundaries.
This open boundary condition ensures 
the energy stability of the system, even when 
strong vortices or backflows
occur at the outflow boundary.
Under certain situations it
can be reduced to a form that can be analogized
to the usual convective boundary condition.
One prominent feature of this boundary condition
is that it provides a control over the velocity
on the outflow/open boundary. This is not available
with the other energy-stable open boundary conditions
from previous works.
Our numerical algorithm treats
the proposed open boundary condition based on
a rotational velocity-correction type strategy.
It gives rise to a Robin-type condition for
the discrete pressure and a Robin-type condition
for the discrete velocity on the outflow/open
boundary, respectively at the pressure and the
velocity sub-steps.
We present extensive numerical experiments 
on a canonical wake flow and a jet flow
in open domain to test the effectiveness and performance
of the method developed herein.
Simulation results are compared with the 
experimental data as well as with other previous simulations
to demonstrate the accuracy of the current method.
Long-time simulations are performed 
for a range of Reynolds numbers, at which 
strong vortices and backflows occur at
the outflow/open boundaries.
The results show that our method is effective in
overcoming the backflow instability,
and that it allows for 
the vortices to discharge from the domain in a fairly
natural fashion even at high Reynolds numbers.

\end{abstract}

%%%%%%%%%%%%%%%%%%%%%%%%%%%%%%%%%%%%%%%%%%%%%%%%%%

\vspace{0.05cm}

%%%
\section{Introduction}
\label{sec:intro}

% what are the issues this paper address?
% what is the status of the field? 
% what has already been done? what has not been done yet?
% what are the convective BC out there?
% what are the traction-free BCs out there?
% what are the effective OBCs in literature?
% 
% what is this paper about?
% what is new about the current OBC?
% what is new about the current algorithm?
%
% e.g. continuing the work in this series DongKC2014, DongS2015
%      we will look into the issue of backflow instability ...
%
% logic:
% (1) what are the issues are you going to address?
%     why is it important?
% (2) what is the status of the field? what effective OBCs
%     are out there?
% (3) what are you going to do about this issue?
%     what is this paper about?
% (4) what are the novelties?
%

The current work focuses on the
outflow/open boundary in incompressible flow simulations
and the issue of
backflow instability, which refers to the
commonly-encountered numerical instability associated with
strong vortices or backflows at the outflow
or open boundaries. 
Extending our efforts on 
this problem~\cite{DongKC2014,Dong2014obc,DongS2015}, 
we strive to develop effective and efficient techniques
to overcome the backflow instability. 

% outflow/open boundaries

A large class of flow problems involve
physically-unbounded domains, such as
jets, wakes, boundary layers, and other
spatially-developing flows.
When numerically simulating such problems,
one will need to truncate the domain artificially
to a finite size and impose some
open (or outflow) boundary condition (OBC) on
the artificial boundary.
Open boundary conditions are among
the most difficult and least understood 
issues in incompressible flow 
simulations~\cite{Gresho1991,SaniG1994},
and have commanded a sustained interest 
of the community for decades.
Among the large volume of works
accumulated so far on this problem,
the traction-free 
condition~\cite{TaylorRM1985,Gartling1990,EngelmanJ1990,
Leone1990,BehrLST1991,SaniG1994,GuermondMS2005,Liu2009}
and the convective (or radiation) boundary 
condition~\cite{Sommerfeld1949,Orlanski1976,Gresho1991,
KeskarL1999,OlshanskiiS2000,ForestierPPS2000,
RuithCM2004,CraskeR2013}
are two of the more commonly used.
We refer the reader to \cite{SaniG1994}
for a review of this field up to the mid-1990s, and
to \cite{PapanastasiouME1992,JinB1993,Johansson1993,HeywoodRT1996,
%Griffiths1997,Renardy1997,
FormaggiaGNQ2002,
HasanAS2005,%NordstromS2005,
NordstromMS2007,
GrinbergK2008,%KimFHJT2009,
PouxGA2011}
%PouxGAA2012} 
for a number of other methods
developed by different researchers.

% backflow instability

Backflow instability is a commonly encountered 
issue with outflows or open boundaries at moderate
and high Reynolds numbers.
Simulations have been observed to instantly 
blow up when strong vortices or backflows occur
at the outflow/open 
boundary~\cite{DongK2005,DongKER2006,VargheseFF2007,GhaisasSF2015}.
As pointed out in~\cite{DongS2015},
a certain amount of backflow at
the outflow boundary appears harmless at low Reynolds numbers,
but when the Reynolds number increases beyond
some moderate value, typically several hundred
to a thousand depending on the geometry 
(e.g. between $Re=300\sim 400$ for the flow past a square cylinder
in two dimensions),
this instability becomes 
a severe issue to numerical simulations.
Commonly-used 
tricks for flow simulations such as
increasing the grid resolution or decreasing
the time step size are observed to 
not help with this instability~\cite{DongKER2006,DongS2015}.

% how do people deal with this instability? 
% large domains? sponge?

For certain
flow problems (e.g. bluff-body wakes)
one way to circumvent this difficulty  is to employ
a large computational domain and to place
the outflow/open boundary far downstream. 
The idea is to allow for the vortices generated 
upstream to sufficiently dissipate before
reaching the outflow boundary.
This is feasible and computationally manageable
at moderate Reynolds numbers.
But this strategy does not scale with
the Reynolds number~\cite{DongKC2014,DongS2015},
because the domain size essential for numerical
stability grows with increasing Reynolds number.
As the Reynolds number becomes large,
the needed domain size for stability can become
very substantial. For example,
in the three-dimensional direct numerical simulation
of the flow past a circular cylinder at
Reynolds number $Re=10000$~\cite{DongKER2006},
a domain size with a wake region $50$ times
the cylinder diameter in length has been used. 
Such a large wake region is essential for
numerical stability for that Reynolds number, even though
the far wake (beyond about $10$ times
the cylinder diameter) is of little or 
no physical interest and the meshes/computations
in that far region are essentially wasted.

% effective OBCs
% what are out there?
% what have already been done?

A far more attractive approach is to devise
effective open/outflow boundary conditions
to overcome the backflow instability.
Several such boundary conditions have been studied
in the literature.
By considering the weak form of the
Navier-Stokes equation and  symmetrization
of the nonlinear term,
Bruneau \& Fabrie~\cite{BruneauF1994,BruneauF1996} 
proposed to modify
the traction condition by a term
$
\frac{1}{2}(\mathbf{n}\cdot\mathbf{u})^-\mathbf{u},
$
where 
$\mathbf{u}$ and $\mathbf{n}$ are respectively
the velocity and the outward-pointing unit vector
normal to the outflow boundary,
and
$
(\mathbf{n}\cdot\mathbf{u})^- 
$
is defined as
$
\mathbf{n}\cdot\mathbf{u}
$
if $\mathbf{n}\cdot\mathbf{u}<0$
and is set to zero otherwise.
We refer to e.g. \cite{LanzendorferS2011,FeistauerN2013}
for applications of this boundary condition
in later works.
%
% Bazilevs et al
A traction condition containing a term
$(\mathbf{n}\cdot\mathbf{u})^-\mathbf{u}$,
which is very similar to that of
\cite{BruneauF1994,BruneauF1996}
but without the $\frac{1}{2}$ factor,
has been employed 
in~\cite{BazilevsGHMZ2009,Moghadametal2011,
PorporaZVP2012,GravemeierCYIW2012,IsmailGCW2014}.
Note that a form 
$\beta(\mathbf{n}\cdot\mathbf{u})^-\mathbf{u}$
where $0<\beta<1$
has also been considered in \cite{Moghadametal2011}.
%
% Dong et al (2014)
By considering the energy balance of
the system, we have proposed in \cite{DongKC2014}
a boundary condition involving a term 
$
\frac{1}{2}|\mathbf{u}|^2\mathbf{n}\Theta_0(\mathbf{n},\mathbf{u}),
$
where $|\mathbf{u}|$ is the velocity
magnitude and
$\Theta_0(\mathbf{n},\mathbf{u})$
is a smoothed step function about 
$\mathbf{n}\cdot\mathbf{u}$
(see Section \ref{sec:method} for definition).
While the role of the term $\Theta_0(\mathbf{n},\mathbf{u})$
can be compared to that of $(\mathbf{n}\cdot\mathbf{u})^-$
discussed above,
the form $\frac{1}{2}|\mathbf{u}|^2\mathbf{n}$
of the OBC in \cite{DongKC2014} is very different from 
those involving $(\mathbf{n}\cdot\mathbf{u})\mathbf{u}$
of the previous 
works~\cite{BruneauF1994,BruneauF1996,BazilevsGHMZ2009,Moghadametal2011,
PorporaZVP2012,GravemeierCYIW2012,IsmailGCW2014}.
%
% velocity penalization
Another boundary condition developed 
in~\cite{BertoglioC2014} employs a penalization of
the tangential velocity derivative
to allow for improved energy balance.
%
% Dong & Shen (2015)
Very recently we have proposed in~\cite{DongS2015} 
a family of open boundary conditions,
having the characteristic that they all
ensure the energy stability of the system
even in situations where strong vortices of
backflows occur at the outflow/open boundary.
This family of boundary conditions contains those 
of \cite{BruneauF1994,BazilevsGHMZ2009,GravemeierCYIW2012,IsmailGCW2014,
DongKC2014} as particular cases,
and more importantly provides other new forms of
open boundary conditions. 
Several of those forms have been investigated in
relative detail in \cite{DongS2015}.

% implementation of OBCs
% difficulties and algorithms
% barriers to adoption

It is observed that, while some of the 
above open boundary conditions
have existed in the literature for some time, 
their adoption in production flow simulations
appears still quite limited. 
This is perhaps in part owing to the challenge
associated with the numerical implementation 
of these boundary conditions. 
All the aforementioned boundary conditions 
for tackling the backflow instability
couple together the velocity
and the pressure, and it is not immediately
clear how to implement them in numerical
simulations.
This seems to be exacerbated by the fact that,
when these boundary conditions are originally proposed,
for most of them their numerical treatments
are not discussed or not adequately discussed,
especially in the context of the commonly-used
splitting or fractional-step type schemes
for incompressible flow simulations.
%This does not help in lowering the barrier
%to the adoption of these boundary conditions
%by the community.
It is noted that in the more recent 
works~\cite{DongKC2014,DongS2015}
two splitting-type schemes,
respectively based on a velocity-correction
type strategy~\cite{DongKC2014}
and a pressure-correction type strategy~\cite{DongS2015},
are presented to deal with
the energy-stable open boundary conditions developed therein.
These algorithms de-couple the computations
for the pressure and the velocity
in the presence of open/outflow boundaries.

% what is this paper about?
% what are the characteristics and advantages of the OBC?

The objective of the current paper is twofold.
First, we present 
a new energy-stable open boundary condition
that is effective in overcoming the backflow instability
for incompressible flow simulations.
This boundary condition involves an inertia
(velocity time-derivative) term, 
and can be shown to ensure the energy stability of
the system even in the presence of
backflows or vortices at the open/outflow
boundary.
It does not belong to the family of
open boundary conditions discussed in~\cite{DongS2015}.
If no backflow occurs at the outflow boundary,
this boundary condition can be
reduced to a form that can be analogized to
the usual convective boundary condition.
Hence, we refer to it as the convective-like
energy-stable open boundary condition. 
The current open boundary condition
has a prominent feature:
it provides a control over the velocity
at the open/outflow boundary.
In contrast, the family of energy-stable open 
boundary conditions from \cite{DongS2015}
and the other 
aforementioned
boundary conditions to address
the backflow instability do not provide
any control over the velocity at
the open/outflow boundary.
Therefore, as the vortices pass through
the outflow/open boundary,
the current
boundary condition can lead to smoother velocity
patterns in regions at or near the outflow
boundary when
compared to that of \cite{DongS2015}.

% algorithm

Second, we present an efficient numerical algorithm 
for treating the proposed open boundary condition. 
Our algorithm overall is based on
a rotational velocity-correction type splitting approach,
and the key issue lies in the numerical treatment of
the inertia term in the open boundary condition.
At the pressure sub-step
our scheme leads to a Robin-type
condition for the discrete pressure on the
outflow boundary, 
and at the velocity sub-step it leads to 
a Robin-type condition for the 
discrete velocity on the outflow boundary.
In contrast, the algorithms of \cite{DongKC2014,DongS2015}
both impose a pressure Dirichlet type condition
on the outflow boundary at the pressure sub-step
and a velocity Neumann type condition
on the outflow boundary at the velocity sub-step.
The current algorithm is  simpler to
implement with spectral-element (and also finite-element) type
spatial discretizations,
because there is no need for the projection of
pressure Dirichlet data on the outflow boundary 
as required by
the algorithms of \cite{DongKC2014,DongS2015}.

% mention that the current OBC can be generalized to a family

We would like to point out that,
by using an idea analogous to that of \cite{DongS2015},
one can generalize the current open boundary
condition to a family of convective-like energy-stable 
open boundary conditions, which will provide other
new forms of OBCs;
see the discussions in Section \ref{sec:obc}
in this regard.
The numerical algorithm presented herein with no change
can be applied together with this family of 
convective-like energy-stable OBCs.

% what are the novelties of this work?

The novelties of this work lie in two aspects:
(i) the convective-like energy-stable open boundary
condition, and (ii) the numerical algorithm for
treating the proposed open boundary condition.
The rotational velocity-correction scheme for 
discretizing the Navier-Stokes equations employed here
has also subtle differences than that 
of~\cite{DongKC2014} in the numerical approximations of
various terms, although both can be classified
as velocity-correction type schemes.

% discuss spatial discretizations

The open boundary condition and the numerical algorithm
developed herein have been implemented and tested
using high-order $C^0$ spectral elements
for spatial 
discretizations~\cite{SherwinK1995,KarniadakisS2005,ZhengD2011}.
The implementations discussed in the paper 
without change can also be
used for finite element discretizations.
It should be noted that the open boundary
condition and the numerical algorithm 
are very general and are not confined to 
a particular spatial discretization technique. They can also
be implemented using other spatial discretization
methods.

% organization of paper

%%%  schemes
\section{Convective-Like Energy-Stable OBC 
and  Algorithm}
\label{sec:method}

\subsection{Convective-Like Energy-Stable Open Boundary Condition}
\label{sec:obc}

% what is the problem formulation?
% what are the governing equations?
% what are the convective energy stable OBCs (CES-OBCs)?
% what are their physical meanings?
% why such forms of OBCs?
% are the OBCs Galilean invariant? do they need to be Galilean invariant?
% what are the initial conditions?

Consider a domain $\Omega$ in two or three dimensions,
and an incompressible flow contained within this domain.
Let $\partial\Omega$ denote the boundary of the domain
$\Omega$. This flow problem is then described by 
the following incompressible Navier-Stokes 
equations in non-dimensional form:
\begin{subequations}
\begin{align}
&
\frac{\partial\mathbf{u}}{\partial t} + \mathbf{u}\cdot\nabla\mathbf{u}
   + \nabla p - \nu\nabla^2\mathbf{u} = \mathbf{f},
\label{equ:nse} \\
&
\nabla\cdot\mathbf{u} = 0,
\label{equ:continuity}
\end{align}
\end{subequations}
where $\mathbf{u}(\mathbf{x},t)$ and $p(\mathbf{x},t)$
are respectively the normalized velocity and pressure fields,
$\mathbf{f}(\mathbf{x},t)$ is an external body force,
and $\mathbf{x}$ and $t$ are the spatial coordinate
and time.
The constant $\nu$ denotes the normalized viscosity,
$\nu=\frac{1}{Re}$, where
$Re$ is the Reynolds number defined by appropriately
choosing a length scale and a velocity scale.

% boundary conditions

We assume that the domain boundary $\partial\Omega$ 
consists of two types:
\begin{equation}
\partial\Omega = \partial\Omega_d \cup \partial\Omega_o,
\quad
\partial\Omega_d \cap \partial\Omega_o = \emptyset.
\end{equation}
In the above $\partial\Omega_d$ denotes the Dirichlet
boundary, on which the velocity is given
\begin{equation}
\mathbf{u} = \mathbf{w}(\mathbf{x},t),
\quad \text{on} \ \partial\Omega_d
\label{equ:dbc}
\end{equation}
where $\mathbf{w}$ is the boundary velocity.
On the other hand, on $\partial\Omega_o$
 neither the velocity $\mathbf{u}$ nor the pressure $p$
is known.
We refer to $\partial\Omega_o$ as the open (or outflow)
boundary. How to deal with $\partial\Omega_o$
in numerical simulations is the focus of the current work.

% CES-OBCs

We propose the following boundary condition for
the open boundary: % $\partial\Omega_o$:
\begin{equation}
 \nu D_0\frac{\partial\mathbf{u}}{\partial t}
-p\mathbf{n} + \nu\mathbf{n}\cdot\nabla\mathbf{u}
  - \frac{1}{2}\left[ 
       \left|\mathbf{u}  \right|^2 \mathbf{n}
       + \left(\mathbf{n} \cdot \mathbf{u} \right)\mathbf{u}
    \right] \Theta_0(\mathbf{n},\mathbf{u})
= \mathbf{f}_b(\mathbf{x},t),
\quad\quad
\text{on } \partial\Omega_o.
\label{equ:obc_D_3}
\end{equation}
In the above equation,  $\mathbf{n}$ is the outward-pointing
unit vector normal to the boundary $\partial\Omega_o$. 
%and $\left|\mathbf{u} \right|$ denotes the magnitude
%of the velocity $\mathbf{u}$.
$D_0$ is a chosen non-negative
constant ($D_0\geqslant 0$), which has been normalized
by $\frac{1}{U_0}$ ($U_0$ denoting the characteristic velocity scale) 
and is non-dimensional.
$\mathbf{f}_b$ is a prescribed vector function on $\partial\Omega_o$
for the purpose of numerical testing only, and
it is set to $\mathbf{f}_b=0$ in actual simulations.
$\Theta_0(\mathbf{n},\mathbf{u})$ is a smoothed
step function about $(\mathbf{n}\cdot\mathbf{u})$
given by \cite{DongKC2014},
\begin{equation}
\Theta_0(\mathbf{n},\mathbf{u}) = \frac{1}{2}\left(
    1 - \tanh \frac{\mathbf{n}\cdot \mathbf{u}}{\delta U_0}
  \right)
\label{equ:Theta_0_expr}
\end{equation}
where $\delta>0$ is a non-dimensional 
positive constant that is 
sufficiently small.
As discussed in \cite{DongKC2014},
$\delta$ controls the sharpness of the smoothed
step function, and it is sharper if $\delta$ is smaller,
and that the simulation result is not sensitive to $\delta$
when it is sufficiently small.
As $\delta \rightarrow 0$, 
$\Theta_0(\mathbf{n},\mathbf{u})$
approaches the step function, that is,
\begin{equation}
\lim_{\delta \rightarrow 0} \Theta_0(\mathbf{n},\mathbf{u}) =
\Theta_{s0}(\mathbf{n},\mathbf{u}) =
\left\{
\begin{array}{ll}
1, & \text{if} \ \mathbf{n}\cdot\mathbf{u}<0, \\
0, & \text{otherwise}.
\end{array}
\right.
\end{equation}

A prominent characteristic of the
open boundary condition \eqref{equ:obc_D_3}
is the inertia term $\frac{\partial\mathbf{u}}{\partial t}$
involved therein.
One can note that for $D_0=0$ the inertia term vanishes and
the boundary 
condition \eqref{equ:obc_D_3} will be reduced
to the so-called boundary condition ``OBC-C''
that has been studied in \cite{DongS2015}.
In the current work we concentrate on
the cases with $D_0>0$.

% generalized form

The open boundary
condition \eqref{equ:obc_D_3},
with $\mathbf{f}_b=0$ and when $\delta$ is sufficiently small,
ensures the energy stability of the system.
To illustrate this point, we look into
the energy balance equation for
the system consisting of
$\eqref{equ:nse}$ and $\eqref{equ:continuity}$:
\begin{equation}
\begin{split}
\frac{\partial}{\partial t} \int_{\Omega} \frac{1}{2}\left|\mathbf{u} \right|^2
=
& -\nu \int_{\Omega}\| \nabla\mathbf{u} \|^2 + \int_{\Omega} \mathbf{f}\cdot \mathbf{u}
  + \int_{\partial\Omega_d} \left(
              \mathbf{n}\cdot\mathbf{T}\cdot \mathbf{u}
              - \frac{1}{2}\left|\mathbf{u} \right|^2 \mathbf{n}\cdot \mathbf{u}
         \right)   \\
&
  + \int_{\partial\Omega_o} \left(
              \mathbf{n}\cdot\mathbf{T} \cdot \mathbf{u}
              - \frac{1}{2}\left|\mathbf{u} \right|^2\mathbf{n}\cdot \mathbf{u}
         \right),
\end{split}
\label{equ:energy}
\end{equation}
where $\mathbf{T} = -p\mathbf{I} + \nu\nabla\mathbf{u}$ 
 and $\mathbf{I}$ denotes the identity tensor.
We assume that $\mathbf{f}_b=0$ in \eqref{equ:obc_D_3} 
and $\delta \rightarrow 0$ in $\Theta_0(\mathbf{n},\mathbf{u})$.
Then by employing the condition \eqref{equ:obc_D_3} on 
$\partial\Omega_o$, the last
surface integral on the right hand side of 
\eqref{equ:energy} becomes
\begin{equation}
\begin{split}
&
\int_{\partial\Omega_o} \left(
              \mathbf{n}\cdot\mathbf{T} \cdot \mathbf{u}
              - \frac{1}{2}\left|\mathbf{u} \right|^2\mathbf{n}\cdot \mathbf{u}
         \right) \\
& = 
-\frac{\partial}{\partial t}\int_{\partial\Omega_o}
     \nu D_0 \frac{1}{2}\left|\mathbf{u}  \right|^2
+ \int_{\partial\Omega_o}
  \frac{1}{2}\left|\mathbf{u}  \right|^2 (\mathbf{n}\cdot\mathbf{u})
       \left[ 2\Theta_{s0}(\mathbf{n},\mathbf{u}) -1 \right]
 \\
&
= -\frac{\partial}{\partial t}\int_{\partial\Omega_o}
     \nu D_0 \frac{1}{2}\left|\mathbf{u}  \right|^2
- \int_{\partial\Omega_o}
  \frac{1}{2}\left|\mathbf{u} \right|^2 \left|\mathbf{n}\cdot\mathbf{u}\right|,
\quad \text{as} \ \delta \rightarrow 0.
%\\
%& \leqslant 
%-\frac{\partial}{\partial t}\int_{\partial\Omega_o}
%     \nu D_0 \frac{1}{2}\left|\mathbf{u}  \right|^2,
%\quad \text{as} \ \delta \rightarrow 0.
\end{split}
\label{equ:surfint_outflow}
\end{equation}
It then follows that  the energy balance equation 
can be transformed into
\begin{equation}
\begin{split}
&
\frac{\partial}{\partial t} \left(
\int_{\Omega} \frac{1}{2}\left|\mathbf{u} \right|^2
+ \nu D_0 \int_{\partial\Omega_o}
      \frac{1}{2}\left|\mathbf{u}  \right|^2
\right) \\
&= 
 -\nu \int_{\Omega}\| \nabla\mathbf{u} \|^2 + \int_{\Omega} \mathbf{f}\cdot \mathbf{u}
  + \int_{\partial\Omega_d} \left(
              \mathbf{n}\cdot\mathbf{T}\cdot \mathbf{u}
              - \frac{1}{2}\left|\mathbf{u} \right|^2 \mathbf{n}\cdot \mathbf{u}
         \right)
- \int_{\partial\Omega_o}
  \frac{1}{2}\left|\mathbf{u} \right|^2 \left|\mathbf{n}\cdot\mathbf{u}\right|,
\quad \text{as} \ \delta \rightarrow 0. 
%+ \int_{\partial\Omega_o}
%  \frac{1}{2}\left|\mathbf{u}  \right|^2 (\mathbf{n}\cdot\mathbf{u})
%       \left[ 2\Theta_{s0}(\mathbf{n},\mathbf{u}) -1 \right] 
%\\
%&
%\leqslant
% -\nu \int_{\Omega}\| \nabla\mathbf{u} \|^2 + \int_{\Omega} \mathbf{f}\cdot \mathbf{u}
%  + \int_{\partial\Omega_d} \left(
%              \mathbf{n}\cdot\mathbf{T}\cdot \mathbf{u}
%              - \frac{1}{2}\left|\mathbf{u} \right|^2 \mathbf{n}\cdot \mathbf{u}
%         \right),
%\quad \text{as} \ \delta \rightarrow 0.   
\end{split}
\label{equ:energy_2}
\end{equation}
Therefore, the open boundary condition given by
\eqref{equ:obc_D_3}, 
when $\mathbf{f}_b=0$ and $\delta$ is sufficiently small,
ensures the energy stability of
the system (in the absence of external forces),
even if strong vortices or backflows occur
(i.e. $\mathbf{n}\cdot\mathbf{u}<0$) 
on the open boundary $\partial\Omega_o$.
Note that, because the velocity $\mathbf{u}$ is given 
on $\partial\Omega_d$, the surface integral term
on $\partial\Omega_d$ in equation
\eqref{equ:energy_2} will not pose a
numerical instability issue.

% comment on control over velocity on open boundary

It is instructive to compare the energy balance
equations  for the current 
open boundary condition and  for the open
boundary conditions
introduced in \cite{DongS2015}.
Let us assume for now that there is no external body
force $\mathbf{f}$ and that
$\mathbf{u}=0$
on the Dirichlet boundary $\partial\Omega_d$.
Then equation \eqref{equ:energy_2} implies that
the sum
$
\left(
\int_{\Omega} \frac{1}{2}\left|\mathbf{u} \right|^2
+ \nu D_0 \int_{\partial\Omega_o}
      \frac{1}{2}\left|\mathbf{u}  \right|^2
\right)
$
will not increase over time.
For $D_0>0$, the energy balance relation
provides an upper bound for
the total energy
$
\int_{\Omega} \frac{1}{2}\left|\mathbf{u} \right|^2
$
and for the quantity
$
\int_{\partial\Omega_o}
      \frac{1}{2}\left|\mathbf{u}  \right|^2
$
with the current open boundary condition.
This provides a control over the velocity $\mathbf{u}$
on the outflow boundary $\partial\Omega_o$.
On the other hand,
with the open boundary conditions from \cite{DongS2015},
the energy balance equation involves only the
total energy, and there is no control
over the velocity on the outflow boundary.
This is a key difference between
the current open boundary condition and
those from \cite{DongS2015}.
Thanks to this characteristic, the current open
boundary condition can lead to qualitatively
smoother velocity patterns at/near
the outflow boundary as vortices pass through.
This point will be illustrated in Section
\ref{sec:tests} using numerical simulations.

% what are the physical meanings of these OBCs?

In addition to the open boundary 
condition \eqref{equ:obc_D_3},
we will also consider the following boundary condition,
\begin{equation}
\nu D_0\frac{\partial\mathbf{u}}{\partial t}
-p\mathbf{n} + \nu\mathbf{n}\cdot\nabla\mathbf{u} = 0,
\quad \text{on} \ \partial\Omega_o,
\label{equ:convec_like_1}
\end{equation}
or equivalently for $D_0> 0$
\begin{equation}
\frac{\partial\mathbf{u}}{\partial t} 
+ \frac{1}{D_0} \frac{\partial\mathbf{u}}{\partial n}
 = \frac{1}{\nu D_0} p\mathbf{n},
\quad \text{on} \ \partial\Omega_o.
\label{equ:convec_like_2}
\end{equation}
The difference between this boundary condition and 
 \eqref{equ:obc_D_3} lies in that
this boundary condition does not 
ensure the energy stability
when backflow occurs on the open boundary $\partial\Omega_o$.
In contrast, the condition \eqref{equ:obc_D_3}
ensures the energy stability even in the presence of
backflows at the open boundary.
Note that for $D_0=0$ the condition 
\eqref{equ:convec_like_1} will be reduced to the
traction-free boundary condition.
One also notes that Equation \eqref{equ:convec_like_2} 
is reminiscent of the usual convective boundary condition
(together with $p=0$),
\begin{equation}
\left\{
\begin{split}
&
\frac{\partial\mathbf{u}}{\partial t} 
+ U_c \frac{\partial\mathbf{u}}{\partial n} = 0, 
\quad \text{on} \ \partial\Omega_o \\
& p = 0, \quad \text{on} \ \partial\Omega_o
\end{split}
\right.
\label{equ:convective_obc}
\end{equation}
where $U_c$ denotes a convection velocity.
Because of this resemblance to the convective boundary condition
we will refer to 
the boundary condition \eqref{equ:obc_D_3}
as a convective-like energy-stable open boundary condition.

% how to choose D_0 in simulations?

The analogy between the current boundary condition and
the usual convective boundary condition suggests that
in the boundary condition \eqref{equ:obc_D_3} 
the parameter $\frac{1}{D_0}$ plays
the role of a convection velocity scale at
the outflow boundary.
Different choices for the convection velocity 
in the usual convective boundary condition
have been considered in a number of studies
(see e.g. \cite{SaniG1994,CraskeR2013}),
which can provide a guide for the choice of $D_0$ in
the boundary condition \eqref{equ:obc_D_3}.
For a given flow problem, one can first perform
a preliminary simulation using the boundary condition
\eqref{equ:obc_D_3} with $D_0=0$ to obtain an estimate
of the convection velocity scale $U_c>0$ at the outflow
boundary, and then carry out the actual simulation
by setting $D_0=\frac{1}{U_c}$ in \eqref{equ:obc_D_3}.
Our numerical experiments in Section \ref{sec:tests}
indicate that the difference in the $D_0$ values
has very little or no effect on
the computed global flow quantities 
such as the force coefficients.
The main effect of $D_0$ appears to be on
the qualitative flow characteristics local to
the outflow boundary.
An improved estimate of the convection velocity
(and hence an improved $D_0$ value in \eqref{equ:obc_D_3})
will allow the vortices or other flow features 
to discharge from
the domain more smoothly and in a more natural fashion.
We will look into the effects of $D_0$ value
in the open boundary condition \eqref{equ:obc_D_3}
in more detail in Section \ref{sec:tests}.

% Remarks on other and general forms

\vspace{0.1in}
{\bf Remarks } \ \
By employing an idea similar to that of \cite{DongS2015},
we can come up with other forms of convective-like 
energy-stable open boundary conditions.
We briefly mention several of them below:
\begin{subequations}
\begin{equation}
 \nu D_0\frac{\partial\mathbf{u}}{\partial t}
-p\mathbf{n} + \nu\mathbf{n}\cdot\nabla\mathbf{u}
  - \left[ 
       \left(\mathbf{n} \cdot \mathbf{u} \right)\mathbf{u}
    \right] \Theta_0(\mathbf{n},\mathbf{u})
= 0, 
\quad\quad
\text{on } \partial\Omega_o;
\label{equ:obc_D_1}
\end{equation}
\begin{equation}
 \nu D_0\frac{\partial\mathbf{u}}{\partial t}
-p\mathbf{n} + \nu\mathbf{n}\cdot\nabla\mathbf{u}
  - \left[ 
       \left|\mathbf{u}  \right|^2 \mathbf{n}
    \right] \Theta_0(\mathbf{n},\mathbf{u})
= 0, 
\quad\quad
\text{on } \partial\Omega_o;
\label{equ:obc_D_2}
\end{equation}
\begin{equation}
 \nu D_0\frac{\partial\mathbf{u}}{\partial t}
-p\mathbf{n} + \nu\mathbf{n}\cdot\nabla\mathbf{u}
  - \left[ \frac{1}{2}
       \left(\mathbf{n} \cdot \mathbf{u} \right)\mathbf{u}
    \right] \Theta_0(\mathbf{n},\mathbf{u})
= 0, 
\quad\quad
\text{on } \partial\Omega_o;
\label{equ:obc_N_1}
\end{equation}
\begin{equation}
 \nu D_0\frac{\partial\mathbf{u}}{\partial t}
-p\mathbf{n} + \nu\mathbf{n}\cdot\nabla\mathbf{u}
  - \left[ 
      \frac{1}{2} \left|\mathbf{u}  \right|^2 \mathbf{n}
    \right] \Theta_0(\mathbf{n},\mathbf{u})
= 0, 
\quad\quad
\text{on } \partial\Omega_o;
\label{equ:obc_N_2}
\end{equation}
\begin{equation}
 \nu D_0\frac{\partial\mathbf{u}}{\partial t}
-p\mathbf{n} + \nu\mathbf{n}\cdot\nabla\mathbf{u}
  - \frac{1}{4}\left[ 
       \left|\mathbf{u}  \right|^2 \mathbf{n}
       + \left(\mathbf{n} \cdot \mathbf{u} \right)\mathbf{u}
    \right] \Theta_0(\mathbf{n},\mathbf{u})
= 0, 
\quad\quad
\text{on } \partial\Omega_o.
\label{equ:obc_N_3}
\end{equation}
\end{subequations}
We would also like to mention the following more 
general form (analogous to \cite{DongS2015}),
which contains the boundary condition \eqref{equ:obc_D_3}
and those represented by \eqref{equ:obc_D_1}--\eqref{equ:obc_N_3}
as particular cases,
\begin{equation}
\nu D_0\frac{\partial\mathbf{u}}{\partial t}
-p\mathbf{n} + \nu\mathbf{n}\cdot\nabla\mathbf{u} 
  - \left[
      \left(\theta + \alpha_2 \right)\frac{1}{2}\left|\mathbf{u} \right|^2 \mathbf{n}  
      + \left(1-\theta + \alpha_1\right) \frac{1}{2}\left(\mathbf{n} \cdot \mathbf{u} \right)\mathbf{u}
    \right] \Theta_0(\mathbf{n},\mathbf{u})
= 0, 
\quad
\text{on } \partial\Omega_o,
\label{equ:obc_gobc}
\end{equation}
where $\theta$, $\alpha_1$ and $\alpha_2$ are constants
satisfying the conditions
\begin{equation}
0 \leqslant \theta \leqslant 1, \quad
\alpha_1 \geqslant 0, \quad
\alpha_2 \geqslant 0.
\end{equation}
Note that the general form \eqref{equ:obc_gobc} 
ensures the energy stability of the system as
$\delta \rightarrow 0$. In this case the energy balance 
relation is given by the following expression,
\begin{equation}
\begin{split}
&
\frac{\partial}{\partial t} \left(
\int_{\Omega} \frac{1}{2}\left|\mathbf{u} \right|^2
+ \nu D_0 \int_{\partial\Omega_o}
      \frac{1}{2}\left|\mathbf{u}  \right|^2
\right) \\
&
= 
-\nu \int_{\Omega}\| \nabla\mathbf{u} \|^2 + \int_{\Omega} \mathbf{f}\cdot \mathbf{u}
  + \int_{\partial\Omega_d} \left(
              \mathbf{n}\cdot\mathbf{T}\cdot \mathbf{u}
              - \frac{1}{2}\left|\mathbf{u} \right|^2 \mathbf{n}\cdot \mathbf{u}
         \right) \\
& \quad
+ \int_{\partial\Omega_o}
     \frac{1}{2}\left|\mathbf{u} \right|^2 \left(\mathbf{n}\cdot \mathbf{u}\right)
     \left[
       \left(1 + \alpha_1+\alpha_2  \right)\Theta_{s0}(\mathbf{n}\cdot\mathbf{u})
       - 1
     \right]
\\
&
\leqslant
 -\nu \int_{\Omega}\| \nabla\mathbf{u} \|^2 + \int_{\Omega} \mathbf{f}\cdot \mathbf{u}
  + \int_{\partial\Omega_d} \left(
              \mathbf{n}\cdot\mathbf{T}\cdot \mathbf{u}
              - \frac{1}{2}\left|\mathbf{u} \right|^2 \mathbf{n}\cdot \mathbf{u}
         \right),
\quad \text{as} \ \delta \rightarrow 0.   
\end{split}
\label{equ:energy_gobc}
\end{equation}

% initial conditions etc

\vspace{0.25in}
Apart from the boundary conditions discussed above,
we impose the following initial condition 
for the velocity,
\begin{equation}
\mathbf{u}(\mathbf{x},t=0) = \mathbf{u}_{in}(\mathbf{x}),
\label{equ:ic}
\end{equation}
where $\mathbf{u}_{in}$ is the initial velocity
field satisfying equation \eqref{equ:continuity}
and compatible with the boundary condition
\eqref{equ:dbc} on $\partial\Omega_d$ at $t=0$.

% what else to discuss here?
% Galilean invariance/variance of boundary conditions?

\subsection{Algorithm Formulation}
\label{sec:algorithm}

The equations \eqref{equ:nse} and \eqref{equ:continuity},
the boundary condition \eqref{equ:dbc} on $\partial\Omega_d$,
and the boundary condition \eqref{equ:obc_D_3}
on $\partial\Omega_o$,
as well as the initial condition \eqref{equ:ic},
together constitute the system that need to be
solved in numerical simulations.

We next present an algorithm for numerically simulating
this system, with emphasis on the numerical
treatment of the open boundary condition \eqref{equ:obc_D_3}.
Let
\begin{equation}
%\mathbf{E}(\mathbf{n},\mathbf{u}) = 
%\left[
%      \left(\theta + \alpha_2 \right)\frac{1}{2}\left|\mathbf{u} \right|^2 \mathbf{n}
%      + \left(1-\theta + \alpha_1\right) \frac{1}{2}\left(\mathbf{n} \cdot \mathbf{u} \right)\mathbf{u}
%    \right] \Theta_0(\mathbf{n},\mathbf{u}),
\mathbf{E}(\mathbf{n},\mathbf{u}) = 
\frac{1}{2}\left[
      \left|\mathbf{u} \right|^2 \mathbf{n}
      +  \left(\mathbf{n} \cdot \mathbf{u} \right)\mathbf{u}
    \right] \Theta_0(\mathbf{n},\mathbf{u}),
\label{equ:E_expr}
\end{equation}
and we re-write equation \eqref{equ:obc_D_3} into 
a more compact form,
\begin{equation}
\nu D_0\frac{\partial\mathbf{u}}{\partial t}
-p\mathbf{n} + \nu\mathbf{n}\cdot\nabla\mathbf{u}
 - \mathbf{E}(\mathbf{n},\mathbf{u}) = \mathbf{f}_b.
\label{equ:obc_gobc_reform}
\end{equation}
We will concentrate on the algorithm and implementation
for $D_0>0$ in \eqref{equ:obc_gobc_reform} 
in this and the next subsections.
In section \ref{sec:D0_zero} we will briefly discuss
how to deal with the case $D_0=0$, when the
current open boundary condition 
is reduced to a form already studied
in \cite{DongS2015}.

Let $n\geqslant 0$ denote the time step index,
and $(\cdot)^n$ denote $(\cdot)$ at time step $n$.
Define $\mathbf{u}^0 = \mathbf{u}_{in}$.
Then, given $\mathbf{u}^n$ we
compute ($p^{n+1}$, $\mathbf{u}^{n+1}$) in a
de-coupled fashion as follows: \\
\underline{For $p^{n+1}$:}
\begin{subequations}
\begin{equation}
\frac{\gamma_0\tilde{\mathbf{u}}^{n+1}-\hat{\mathbf{u}}}{\Delta t}
+ \mathbf{u}^{*,n+1}\cdot\nabla\mathbf{u}^{*,n+1}
+ \nabla p^{n+1}
+ \nu \nabla\times\nabla\times\mathbf{u}^{*,n+1}
= \mathbf{f}^{n+1}
\label{equ:pressure_1}
\end{equation}
\begin{equation}
\nabla\cdot\tilde{\mathbf{u}}^{n+1} = 0
\label{equ:pressure_2}
\end{equation}
\begin{equation}
\mathbf{n}\cdot\tilde{\mathbf{u}}^{n+1} 
 = \mathbf{n} \cdot \mathbf{w}^{n+1},
\quad \text{on} \ \partial\Omega_d
\label{equ:pressure_3}
\end{equation}
\begin{equation}
\nu D_0\frac{\gamma_0\tilde{\mathbf{u}}^{n+1}-\hat{\mathbf{u}}}{\Delta t}\cdot\mathbf{n}
- p^{n+1}
+ \nu\mathbf{n}\cdot\nabla\mathbf{u}^{*,n+1}\cdot\mathbf{n}
- \mathbf{n}\cdot\mathbf{E}(\mathbf{n},\mathbf{u}^{*,n+1})
= \mathbf{f}_b^{n+1}\cdot\mathbf{n},
\quad \text{on} \ \partial\Omega_o
\label{equ:pressure_4}
\end{equation}
\end{subequations}
\underline{For $\mathbf{u}^{n+1}$:}
\begin{subequations}
\begin{equation}
\frac{\gamma_0\mathbf{u}^{n+1}-\gamma_0\tilde{\mathbf{u}}^{n+1}}{\Delta t}
- \nu\nabla^2\mathbf{u}^{n+1} 
= \nu \nabla\times\nabla\times\mathbf{u}^{*,n+1}
\label{equ:velocity_1}
\end{equation}
\begin{equation}
\mathbf{u}^{n+1} = \mathbf{w}^{n+1},
\quad \text{on} \ \partial\Omega_d
\label{equ:velocity_2}
\end{equation}
\begin{equation}
\nu D_0\frac{\gamma_0\mathbf{u}^{n+1}-\hat{\mathbf{u}}}{\Delta t}
- p^{n+1}\mathbf{n} 
+ \nu\mathbf{n}\cdot\nabla\mathbf{u}^{n+1}
- \mathbf{E}(\mathbf{n},\mathbf{u}^{*,n+1})
+ \nu\left(\nabla\cdot\mathbf{u}^{*,n+1}  \right)\mathbf{n}
= \mathbf{f}_b^{n+1},
\quad \text{on} \ \partial\Omega_o.
\label{equ:velocity_3}
\end{equation}
\end{subequations}
%
% meanings of symbols
%
In the above equations, 
$\Delta t$ is the time step size,
$\mathbf{n}$ is the outward-pointing unit vector normal
to the boundary,
and $\tilde{\mathbf{u}}^{n+1}$ is an auxiliary variable
approximating $\mathbf{u}^{n+1}$. 
Let $J$ ($J=1$ or $2$) denote the
temporal order of accuracy of the algorithm.
Then $\mathbf{u}^{*,n+1}$ is a $J$-th order 
explicit approximation of $\mathbf{u}^{n+1}$ given by
\begin{equation}
\mathbf{u}^{*,n+1} = \left\{
\begin{array}{ll}
\mathbf{u}^n, & J=1, \\
2\mathbf{u}^n - \mathbf{u}^{n-1}, & J=2.
\end{array}
\right.
\end{equation}
The expressions 
$
\frac{1}{\Delta t}(\gamma_0\mathbf{u}^{n+1}-\hat{\mathbf{u}})
$
and 
$
\frac{1}{\Delta t}(\gamma_0\tilde{\mathbf{u}}^{n+1}-\hat{\mathbf{u}})
$
are approximations of 
$
\left. 
\frac{\partial\mathbf{u}}{\partial t}
\right|^{n+1}
$
by a $J$-th order backward differentiation
formula, and  $\hat{\mathbf{u}}$ and $\gamma_0$
are given by
\begin{equation}
\hat{\mathbf{u}} = \left\{
\begin{array}{ll}
\mathbf{u}^n, & J=1, \\
2\mathbf{u}^n-\frac{1}{2}\mathbf{u}^{n-1}, & J=2,
\end{array}
\right.
\qquad\quad
\gamma_0 = \left\{
\begin{array}{ll}
1, & J=1, \\
\frac{3}{2}, & J=2.
\end{array}
\right.
\end{equation}
Note that $\mathbf{E}(\mathbf{n},\mathbf{u})$
is given by \eqref{equ:E_expr}.

% comments on the algorithm
% compare with DongKC2014 scheme, DongS2010, DongS2015

One can observe that the overall structure of
the above algorithm represents a rotational velocity-correction
type strategy (see \cite{GuermondS2003a,DongS2010,DongS2012,DongKC2014})
for de-coupling the computations
of the pressure and velocity. 
While both belong to velocity correction-type schemes,
the scheme here is somewhat different from the one of \cite{DongKC2014}.
Note that in the pressure sub-step we have approximated all terms
at time step $(n+1)$ with the current 
scheme (see equation \eqref{equ:pressure_1}). 
In contrast,
in \cite{DongKC2014} the viscous and the nonlinear
terms are approximated at time step $n$ 
rather than at $(n+1)$ in the pressure sub-step,
and correspondingly some correction terms are
incorporated in the subsequent velocity sub-step.
The current treatment of various terms
is observed to yield smaller pressure errors and comparable
velocity errors compared to that of \cite{DongKC2014}.

The inertia term $\nu D_0\frac{\partial\mathbf{u}}{\partial t}$ in
the boundary condition \eqref{equ:obc_gobc_reform}
demands some care in the temporal discretization.
%This distinguishes
%the current algorithm from that of \cite{DongKC2014},
%while both are of rotational velocity-correction types.
The discrete equation \eqref{equ:pressure_4} 
in the pressure sub-step 
stems from an inner product between
the directional vector $\mathbf{n}$ and
the open boundary condition \eqref{equ:obc_gobc_reform}
on the outflow boundary $\partial\Omega_o$.
Note that the $\frac{\partial\mathbf{u}}{\partial t}$ term
and the pressure term have been treated implicitly 
 in \eqref{equ:pressure_4}, and in particular that 
$\frac{\partial\mathbf{u}}{\partial t}$ is approximated
using $\tilde{\mathbf{u}}^{n+1}$ (instead of
$\mathbf{u}^{n+1}$) here in this discrete equation.
This point is crucial, and it effectively leads to
a Robin-type condition for the pressure $p^{n+1}$
on $\partial\Omega_o$
because of the equation \eqref{equ:pressure_1}.
An explicit treatment of 
the $\frac{\partial\mathbf{u}}{\partial t}$ term
in \eqref{equ:pressure_4} would seem more attractive and 
would result in
a Dirichlet type condition for $p^{n+1}$
on $\partial\Omega_o$, just like in the scheme of
 \cite{DongKC2014}.
This treatment however does not work, and is unstable
unless $D_0$ is very small.
At the velocity sub-step,
in the discrete equation \eqref{equ:velocity_3} on $\partial\Omega_o$
we have treated the 
terms $\frac{\partial\mathbf{u}}{\partial t}$ 
and $\mathbf{n}\cdot\nabla\mathbf{u}$ implicitly,
and note that
$\frac{\partial\mathbf{u}}{\partial t}$ is approximated
using $\mathbf{u}^{n+1}$ here.
These numerical treatments give rise to a Robin-type 
condition for the discrete velocity $\mathbf{u}^{n+1}$
on $\partial\Omega_o$, noting that in \eqref{equ:velocity_3}
$p^{n+1}$ is already explicitly known from
the pressure sub-step.
Note also that in the discrete equation
\eqref{equ:velocity_3}  an extra term
$\nu (\nabla\cdot\mathbf{u}) \mathbf{n}$
has been incorporated in the formulation.

% what else to discuss here?

We would like to point out that the algorithmic
formulation given
by \eqref{equ:pressure_1}--\eqref{equ:velocity_3}
can be used together with the general form of 
convective-like energy-stable open
boundary condition \eqref{equ:obc_gobc},
by setting $\mathbf{E}(\mathbf{n},\mathbf{u})$ in
the algorithm
as follows
\begin{equation}
\mathbf{E}(\mathbf{n},\mathbf{u}) = 
\left[
      \left(\theta + \alpha_2 \right)\frac{1}{2}\left|\mathbf{u} \right|^2 \mathbf{n}
      + \left(1-\theta + \alpha_1\right) \frac{1}{2}\left(\mathbf{n} \cdot \mathbf{u} \right)\mathbf{u}
    \right] \Theta_0(\mathbf{n},\mathbf{u}).
\label{equ:E_expr_2}
\end{equation}

\subsection{Implementation with $C^0$ Spectral Elements}
\label{sec:implementation}

We employ $C^0$-continuous high-order spectral elements 
\cite{SherwinK1995,KarniadakisS2005,
ZhengD2011}
for spatial discretizations in the current work.
Let us next look into how to implement the algorithm,
given by \eqref{equ:pressure_1}--\eqref{equ:velocity_3},
using $C^0$  spectral elements.
The discussions in this subsection with no change
also apply to $C^0$ finite element implementations.

As noted in several previous 
works~\cite{DongS2010,DongS2012,Dong2012,DongKC2014,Dong2014nphase},
the complication in the implementation
with $C^0$ elements stems from the high-order derivative
terms such as $\nabla\times\nabla\times\mathbf{u}$
involved in this type of algorithm,
because such terms cannot be directly
computed in the discrete function
space of $C^0$ elements.
We can eliminate such complications by looking
into the weak forms of the algorithm.
In addition, we will  eliminate 
the auxiliary velocity $\tilde{\mathbf{u}}^{n+1}$
from the final form of the algorithm.

We first formulate the weak forms of the algorithm in
the spatially continuous space.
Let $q(\mathbf{x})$ denote a test function.
By taking the $L^2$ inner product between $\nabla q$
and equation \eqref{equ:pressure_1} and integrating by part,
we have
\begin{equation}
\begin{split}
\int_{\Omega} \nabla p^{n+1}\cdot \nabla q
+ \frac{\gamma_0}{\Delta t} \int_{\partial\Omega_o}
     \mathbf{n}\cdot \tilde{\mathbf{u}}^{n+1} q
= & \int_{\Omega} \mathbf{G}^{n+1}\cdot\nabla q
- \nu\int_{\partial\Omega_d\cup\partial\Omega_o} 
     \mathbf{n}\times \bm{\omega}^{*,n+1}\cdot\nabla q \\
&
- \frac{\gamma_0}{\Delta t}\int_{\partial\Omega_d}
     \mathbf{n}\cdot \mathbf{w}^{n+1} q,
\qquad \forall q,
\end{split}
\label{equ:p_weak_1}
\end{equation}
where $\bm{\omega}=\nabla\times\mathbf{u}$ is the vorticity,
\begin{equation}
\mathbf{G}^{n+1} = \mathbf{f}^{n+1} + \frac{\hat{\mathbf{u}}}{\Delta t}
- \mathbf{u}^{*,n+1}\cdot\nabla\mathbf{u}^{*,n+1},
\label{equ:G_expr}
\end{equation}
and we have used equations \eqref{equ:pressure_2} and
\eqref{equ:pressure_3}, the divergence theorem,
and the identify
\begin{equation}
\int_{\Omega}\nabla\times\bm{\omega} \cdot\nabla q
= \int_{\Omega} \nabla\cdot(\bm{\omega}\times\nabla q)
= \int_{\partial\Omega}\mathbf{n}\times\bm{\omega}\cdot\nabla q.
\end{equation}
According to equation \eqref{equ:pressure_4},
$\mathbf{n}\cdot\tilde{\mathbf{u}}^{n+1}$ can be expressed
in terms of $p^{n+1}$ and other explicit quantities
on $\partial\Omega_o$. We therefore can 
transform equation \eqref{equ:p_weak_1}
into the final weak form for the pressure $p^{n+1}$,
\begin{equation}
\begin{split}
\int_{\Omega} \nabla p^{n+1}\cdot \nabla q
&+ \frac{1}{\nu D_0} \int_{\partial\Omega_o}
     p^{n+1} q
=  \int_{\Omega} \mathbf{G}^{n+1}\cdot\nabla q
- \nu\int_{\partial\Omega_d\cup\partial\Omega_o} 
     \mathbf{n}\times \bm{\omega}^{*,n+1}\cdot\nabla q \\
&
+ \int_{\partial\Omega_o} \left\{
  -\frac{1}{\Delta t}\mathbf{n}\cdot\hat{\mathbf{u}}
  + \frac{1}{\nu D_0}\left[
    \nu\mathbf{n}\cdot\nabla\mathbf{u}^{*,n+1}\cdot\mathbf{n}
     - \mathbf{n}\cdot\mathbf{E}(\mathbf{n},\mathbf{u}^{*,n+1})
     - \mathbf{f}_b^{n+1}\cdot\mathbf{n}
  \right]
\right\} q \\
&
- \frac{\gamma_0}{\Delta t}\int_{\partial\Omega_d}
     \mathbf{n}\cdot \mathbf{w}^{n+1} q,
\qquad \forall q.
\end{split}
\label{equ:p_weakform}
\end{equation}

We next sum up equations \eqref{equ:velocity_1} 
and \eqref{equ:pressure_1}
to obtain
\begin{equation}
\frac{\gamma_0}{\nu \Delta t}\mathbf{u}^{n+1}
 - \nabla^2\mathbf{u}^{n+1}
= \frac{1}{\nu}\left(
  \mathbf{G}^{n+1} - \nabla p^{n+1}
\right).
\label{equ:vel_helm}
\end{equation}
Let $\varphi(\mathbf{x})$ denote a test function 
that vanishes on $\partial\Omega_d$.
Taking the $L^2$ inner product between $\varphi$
and equation \eqref{equ:vel_helm} and integrating
by part lead to 
\begin{equation}
\frac{\gamma_0}{\nu\Delta t} \int_{\Omega}\mathbf{u}^{n+1}\varphi
+ \int_{\Omega}\nabla\varphi\cdot \nabla\mathbf{u}^{n+1}
- \int_{\partial\Omega_o} \mathbf{n}\cdot\nabla\mathbf{u}^{n+1}\varphi
= \frac{1}{\nu}\int_{\Omega}\left(
  \mathbf{G}^{n+1}-\nabla p^{n+1}
\right)\varphi,
\qquad \forall \varphi,
\label{equ:u_weak_1}
\end{equation}
where we have used the divergence theorem,
and the fact that 
$
\int_{\partial\Omega_d} \mathbf{n}\cdot\nabla\mathbf{u}^{n+1}\varphi =0
$
thanks to the requirement that 
$\varphi=0$ on $\partial\Omega_d$.
According to equation \eqref{equ:velocity_3},
$\mathbf{n}\cdot\nabla\mathbf{u}^{n+1}$
can be expressed in terms of $\mathbf{u}^{n+1}$ and
other explicit quantities on $\partial\Omega_o$. 
We therefore can transform \eqref{equ:u_weak_1}
into the final weak form for $\mathbf{u}^{n+1}$,
\begin{equation}
\begin{split}
\frac{\gamma_0}{\nu\Delta t} \int_{\Omega}\mathbf{u}^{n+1}\varphi
&+ \int_{\Omega}\nabla\varphi\cdot\nabla\mathbf{u}^{n+1}
+ \frac{\gamma_0 D_0}{\Delta t} \int_{\partial\Omega_o} 
       \mathbf{u}^{n+1}\varphi
= \frac{1}{\nu}\int_{\Omega}\left(
  \mathbf{G}^{n+1}-\nabla p^{n+1}
\right)\varphi \\
&
+ \int_{\partial\Omega_o}\left\{
  \frac{D_0}{\Delta t}\hat{\mathbf{u}}
  + \frac{1}{\nu}\left[
    p^{n+1}\mathbf{n} + \mathbf{E}(\mathbf{n},\mathbf{u}^{*,n+1})
    + \mathbf{f}_b^{n+1}
    - \nu \left(\nabla\cdot\mathbf{u}^{*,n+1}  \right)\mathbf{n}
  \right]
\right\} \varphi,
\quad \forall \varphi.
\end{split}
\label{equ:u_weakform}
\end{equation}

The weak forms of the algorithm in the continuum space
 consist of equations
\eqref{equ:p_weakform} and \eqref{equ:u_weakform},
together with the velocity Dirichlet condition
\eqref{equ:velocity_2} on $\partial\Omega_d$.
The auxiliary variable $\tilde{\mathbf{u}}^{n+1}$
does not appear in the weak form and is not explicitly
computed.
These equations  in weak forms  can be
discretized using $C^0$  spectral
elements (or finite elements) in a straightforward fashion.

% spatial discretizations
% restrict the space of q and varphi

Let $\Omega_h$ denote the domain $\Omega$ discretized using
a spectral element mesh, 
and $\partial\Omega_h = \partial\Omega_{dh}\cup\partial\Omega_{oh}$
denote the boundary of $\Omega_h$,
where $\partial\Omega_{dh}$ and $\partial\Omega_{oh}$
respectively represent the discretized $\partial\Omega_d$
and $\partial\Omega_o$.
Let $\mathbf{X}_h\subset [H^1(\Omega_h)]^d$ 
(where $d=2$ or $3$ is the spatial
dimension) denote the approximation space
of the discretized velocity $\mathbf{u}_h^{n+1}$,
and define 
$
X_{h0} = \left\{ \
  v \in H^1(\Omega_h) \ :\ v|_{\partial\Omega_{dh}} = 0
\ \right\}.
$
Let $M_h \subset H^1(\Omega_h)$
denote the approximation space of 
the discretized pressure $p^{n+1}$.
We take the test function $q$ of 
equation \eqref{equ:p_weakform} from 
$M_h$, and take the test function $\varphi$
of equation \eqref{equ:u_weakform} from $X_{h0}$.
In the following let $(\cdot)_h$  denote
the discretized version of the variable $(\cdot)$.
Then the discretized
version of equation \eqref{equ:p_weakform}
is: Find $p_h^{n+1}\in M_h$ such that
\begin{equation}
\begin{split}
\int_{\Omega_h} & \nabla p_h^{n+1}\cdot \nabla q_h
+ \frac{1}{\nu D_0} \int_{\partial\Omega_{oh}}
     p_h^{n+1} q_h
=  \int_{\Omega_h} \mathbf{G}_h^{n+1}\cdot\nabla q_h
- \nu\int_{\partial\Omega_{dh}\cup\partial\Omega_{oh}} 
     \mathbf{n}_h\times \bm{\omega}_h^{*,n+1}\cdot\nabla q_h \\
&
+ \int_{\partial\Omega_{oh}} \left\{
  -\frac{1}{\Delta t}\mathbf{n}_h\cdot\hat{\mathbf{u}}_h
  + \frac{1}{\nu D_0}\left[
    \nu\mathbf{n}_h\cdot\nabla\mathbf{u}_h^{*,n+1}\cdot\mathbf{n}_h
     - \mathbf{n}_h\cdot\mathbf{E}(\mathbf{n}_h,\mathbf{u}_h^{*,n+1})
     - \mathbf{f}_{bh}^{n+1}\cdot\mathbf{n}_h
  \right]
\right\} q_h \\
&
- \frac{\gamma_0}{\Delta t}\int_{\partial\Omega_{dh}}
     \mathbf{n}_h\cdot \mathbf{w}_h^{n+1} q_h,
\qquad \forall q_h \in M_h.
\end{split}
\label{equ:p_weakform_disc}
\end{equation}
The discretized version of equations \eqref{equ:u_weakform}
and \eqref{equ:velocity_2}
is: Find $\mathbf{u}_h^{n+1}\in \mathbf{X}_h$ such that
\begin{equation}
\begin{split}
\frac{\gamma_0}{\nu\Delta t} \int_{\Omega_h} & \mathbf{u}_h^{n+1}\varphi_h
+ \int_{\Omega_h}\nabla\varphi_h\cdot\nabla\mathbf{u}_h^{n+1}
+ \frac{\gamma_0 D_0}{\Delta t} \int_{\partial\Omega_{oh}} 
       \mathbf{u}_h^{n+1}\varphi_h
= \frac{1}{\nu}\int_{\Omega_h}\left(
  \mathbf{G}_h^{n+1}-\nabla p_h^{n+1}
\right)\varphi_h \\
&
+ \int_{\partial\Omega_{oh}}\left\{
  \frac{D_0}{\Delta t}\hat{\mathbf{u}}_h
  + \frac{1}{\nu}\left[
    p_h^{n+1}\mathbf{n}_h + \mathbf{E}(\mathbf{n}_h,\mathbf{u}_h^{*,n+1})
    + \mathbf{f}_{bh}^{n+1}
    - \nu \left(\nabla\cdot\mathbf{u}_h^{*,n+1}  \right)\mathbf{n}_h
  \right]
\right\} \varphi_h, \\
&
\quad \forall \varphi_h\in X_{h0},
\end{split}
\label{equ:u_weakform_disc}
\end{equation}
together with
\begin{equation}
\mathbf{u}_h^{n+1} = \mathbf{w}_h^{n+1},
\qquad \text{on} \ \partial\Omega_{dh}.
\label{equ:dbc_disc}
\end{equation}

% comment on final algorithm, characteristics etc

Our final algorithm therefore consists of the following
operations within a time step:
(i) Solve equation \eqref{equ:p_weakform_disc}
for $\mathbf{p}_h^{n+1}$;
(ii) Solve equation \eqref{equ:u_weakform_disc},
together with the Dirichlet condition \eqref{equ:dbc_disc}
on $\partial\Omega_{dh}$, for $\mathbf{u}_h^{n+1}$.
The computations for the pressure and the velocity are
de-coupled, and the computations for the three
components of the velocity are also de-coupled.
All the terms on the right hand sides of
equations \eqref{equ:p_weakform_disc} and
\eqref{equ:u_weakform_disc} can be
computed directly using $C^0$ spectral elements.
Note that the auxiliary velocity $\tilde{\mathbf{u}}^{n+1}$
is not explicitly computed.

% equal order etc

We employ equal orders of expansion polynomials
to approximate the pressure and the velocity in 
the current spectral-element implementation,
similar to our previous 
works~\cite{DongS2010,DongS2012,Dong2012,
Dong2014obc,DongKC2014,DongS2015}.
Note that in all the numerical simulations and flow 
tests of Section \ref{sec:tests} we have
used the same polynomial orders 
for the pressure and the velocity.
We refer the reader to the
equal-order approximations for the pressure/velocity
by other researchers in the 
literature~\cite{KarniadakisIO1991,TimmermansMV1996,
GuermondS2003,KarniadakisS2005,LiuLP2007,
Liu2009,BazilevsGHMZ2009,Moghadametal2011}.

% what else to discuss here?

\subsection{The Case of $D_0=0$ in Open Boundary Condition}
\label{sec:D0_zero}

%In Sections \ref{sec:algorithm} and \ref{sec:implementation}
So far we have focused on the case
 $D_0>0$ in the open boundary 
condition \eqref{equ:obc_gobc_reform}.
In this subsection we briefly discuss the case
$D_0=0$ in the boundary condition. 

As noted in Section \ref{sec:obc},
with $D_0=0$ the boundary condition \eqref{equ:obc_D_3}
is reduced to a form (so-called ``OBC-C'') that is already studied 
in \cite{DongS2015}.
One can therefore employ the algorithms from
\cite{DongS2015} or \cite{DongKC2014} to treat
this case. Note that the algorithm presented
in \cite{DongKC2014} is with respect to 
the open boundary condition 
having a form corresponding
to the so-called ``OBC-E'' in \cite{DongS2015}.
%with a form corresponding to
%$
%\mathbf{E}(\mathbf{n},\mathbf{u}) = 
%\frac{1}{2}|\mathbf{u}|^2\mathbf{n}\Theta_0(\mathbf{n},\mathbf{u}).
%$
%corresponding to $(\theta,\alpha_1,\alpha_2)=(1,0,0)$,
But the algorithm of \cite{DongKC2014}
also applies to other forms of open boundary conditions
given in \cite{DongS2015}.

With $D_0=0$
the essential difference when 
compared with the scheme presented
in Section \ref{sec:algorithm} lies in that, 
in the pressure sub-step the pressure
condition on the open boundary will now become of
Dirichlet type rather than Robin type, 
and in the velocity sub-step the velocity condition
on the open boundary will become of 
Neumann type rather than Robin type. 

We now briefly mention an algorithm for $D_0=0$,
as an alternative to the one presented in \cite{DongKC2014}.
We discretize the governing equations and the boundary
conditions as follows: \\
\underline{For $p^{n+1}$:}
\begin{subequations}
\begin{equation*}
\text{
Use equations \eqref{equ:pressure_1}, \eqref{equ:pressure_2},
and \eqref{equ:pressure_3};
}
\end{equation*}
\begin{equation}
 p^{n+1} =
 \nu\mathbf{n}\cdot\nabla\mathbf{u}^{*,n+1}\cdot\mathbf{n}
- \mathbf{n}\cdot\mathbf{E}(\mathbf{n},\mathbf{u}^{*,n+1})
- \mathbf{f}_b^{n+1}\cdot\mathbf{n},
\quad \text{on} \ \partial\Omega_o.
\label{equ:D0_zero_pressure_4}
\end{equation}
\end{subequations}
\underline{For $\mathbf{u}^{n+1}$:}
\begin{subequations}
\begin{equation*}
\text{
Use equations \eqref{equ:velocity_1} and \eqref{equ:velocity_2};
}
\end{equation*}
\begin{equation}
\mathbf{n}\cdot\nabla\mathbf{u}^{n+1} =
\frac{1}{\nu}\left[
  p^{n+1}\mathbf{n} 
  + \mathbf{E}(\mathbf{n},\mathbf{u}^{*,n+1})
  - \nu\left(\nabla\cdot\mathbf{u}^{*,n+1}  \right)\mathbf{n}
  + \mathbf{f}_b^{n+1}
\right],
\quad \text{on} \ \partial\Omega_o.
\label{equ:D0_zero_velocity_3}
\end{equation}
\end{subequations}
The difference between this algorithm and that of \cite{DongKC2014}
lies in that, in the pressure sub-step of this algorithm 
we have approximated
all terms at the time step $(n+1)$ and in the velocity sub-step
no correction terms are involved.
On the other hand, in \cite{DongKC2014}
certain terms are approximated at time step $(n+1)$ and the other
terms are approximated at step $n$ in the pressure sub-step,
and in the velocity sub-step several correction terms
are involved as a result.

The weak forms of this algorithm can be obtained
using a procedure similar to that of \cite{DongKC2014}.
Let 
$
H_{p0}^1(\Omega) = \left\{ \
  v\in H^1(\Omega) \ : \
  v|_{\partial\Omega_o} = 0
\ \right\},
$
and $q \in H_{p0}(\Omega)$ denote
the test function.
Then the weak form for $p^{n+1}$ is
\begin{equation}
\begin{split}
\int_{\Omega} \nabla p^{n+1}\cdot \nabla q
= & \int_{\Omega} \mathbf{G}^{n+1}\cdot\nabla q
- \nu\int_{\partial\Omega_d\cup\partial\Omega_o} 
     \mathbf{n}\times \bm{\omega}^{*,n+1}\cdot\nabla q \\
&
- \frac{\gamma_0}{\Delta t}\int_{\partial\Omega_d}
     \mathbf{n}\cdot \mathbf{w}^{n+1} q,
\qquad \forall q\in H_{p0}^1(\Omega),
\end{split}
\label{equ:D0_zero_p_weakform}
\end{equation}
where $\mathbf{G}^{n+1}$ is given by \eqref{equ:G_expr}.
Let 
$
H_{u0}^1(\Omega) = \left\{ \
  v \in H^1(\Omega) \ : \
  v|_{\partial\Omega_d} = 0
\ \right\},
$
and $\varphi \in H_{u0}^1(\Omega)$ denote the test function.
Then the weak form for $\mathbf{u}^{n+1}$ is 
\begin{equation}
\begin{split}
\frac{\gamma_0}{\nu\Delta t} \int_{\Omega}\mathbf{u}^{n+1}\varphi
&+ \int_{\Omega}\nabla\varphi\cdot\nabla\mathbf{u}^{n+1}
= \frac{1}{\nu}\int_{\Omega}\left(
  \mathbf{G}^{n+1}-\nabla p^{n+1}
\right)\varphi \\
&
+ \frac{1}{\nu} \int_{\partial\Omega_o}
   \left[
    p^{n+1}\mathbf{n} + \mathbf{E}(\mathbf{n},\mathbf{u}^{*,n+1})
    + \mathbf{f}_b^{n+1}
    - \nu \left(\nabla\cdot\mathbf{u}^{*,n+1}  \right)\mathbf{n}
  \right] \varphi,
\quad \forall \varphi \in H_{u0}^1(\Omega).
\end{split}
\label{equ:D0_zero_u_weakform}
\end{equation}

The algorithm involves the following operations within
a time step:
(i) Solve equation \eqref{equ:D0_zero_p_weakform}, together
with the pressure Dirichlet condition \eqref{equ:D0_zero_pressure_4}
on $\partial\Omega_o$, for $p^{n+1}$;
(ii) Solve equation \eqref{equ:D0_zero_u_weakform}, together with
the velocity Dirichlet condition \eqref{equ:velocity_2}
on $\partial\Omega_d$, for $\mathbf{u}^{n+1}$.
When imposing the pressure Dirichlet 
condition \eqref{equ:D0_zero_pressure_4}
on $\partial\Omega_o$ using $C^0$ spectral elements (or
finite elements),
a projection of the pressure Dirichlet data to 
the $H^1(\partial\Omega_o)$ space is required
because of the velocity gradient term involved in 
the equation; see \cite{DongKC2014} for more detailed
discussions in this regard.
We have implemented the above algorithm,
and the numerical experiments reported
 in Section \ref{sec:tests}
corresponding to $D_0=0$
 are performed using
this algorithm.

% what else to discuss here?

%%% Test problems: 
\section{Representative Numerical Tests}
\label{sec:tests}

In this section we consider several
flow problems with open/outflow boundaries
and employ two-dimensional simulations
to demonstrate
the effectiveness and performance of the
open boundary condition and the numerical
algorithm developed in the previous 
section.
At large Reynolds numbers
the presence of vortices and backflows
at the open/outflow boundary 
makes these problems very challenging to simulate.
We will look into the spatial and temporal
convergence rates of the algorithm,
and compare current simulation results
with the experimental data as well as other
simulations from the literature.
The results show the effectiveness
of the proposed method for dealing with
the backflow instability.

\subsection{Convergence Rates}

% analytic solution, spatial/temporal convergence

\begin{figure}
\centerline{
\includegraphics[width=3.5in]{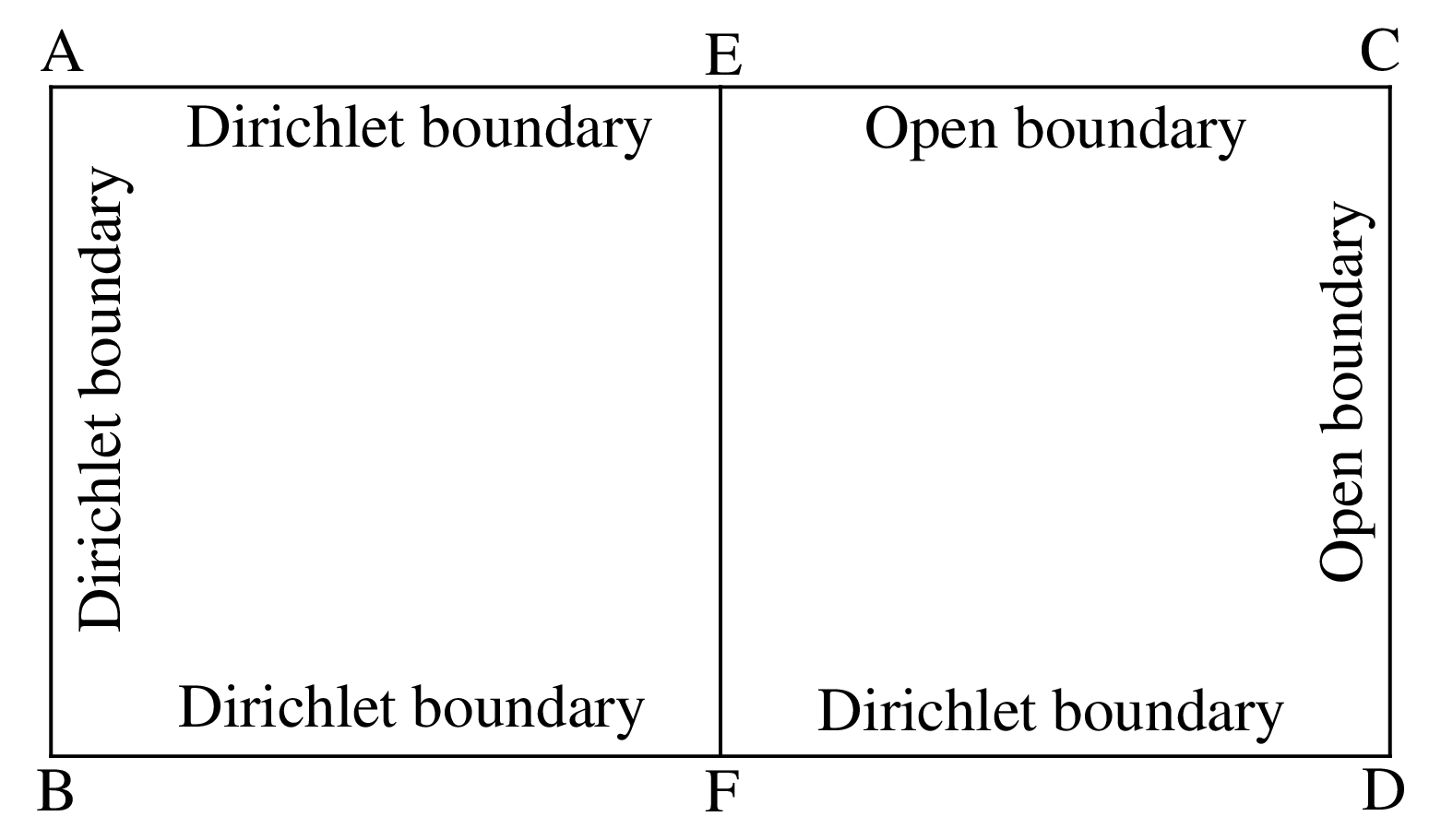}(a)
}
\centerline{
\includegraphics[width=3in]{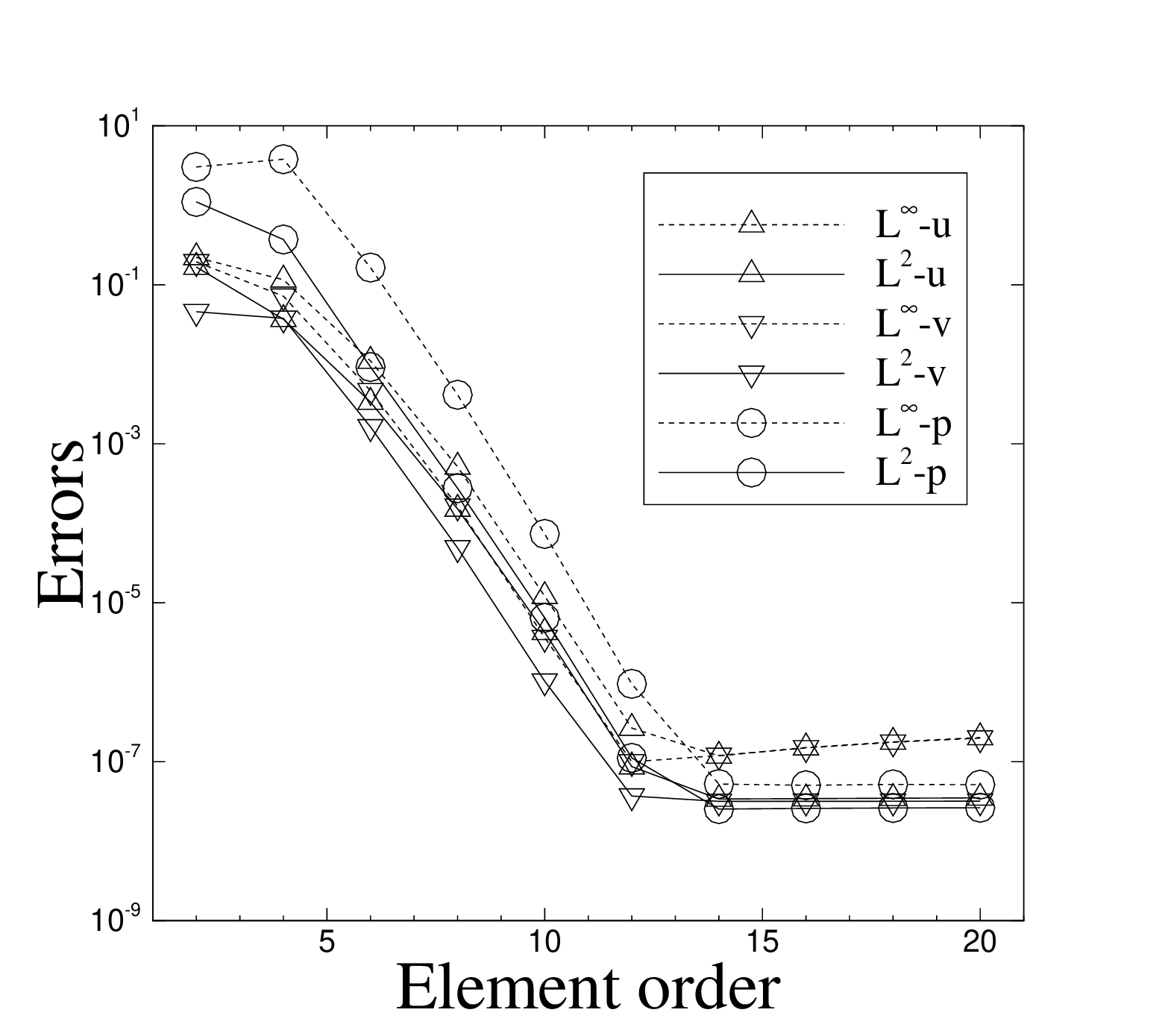}(b)
\includegraphics[width=3in]{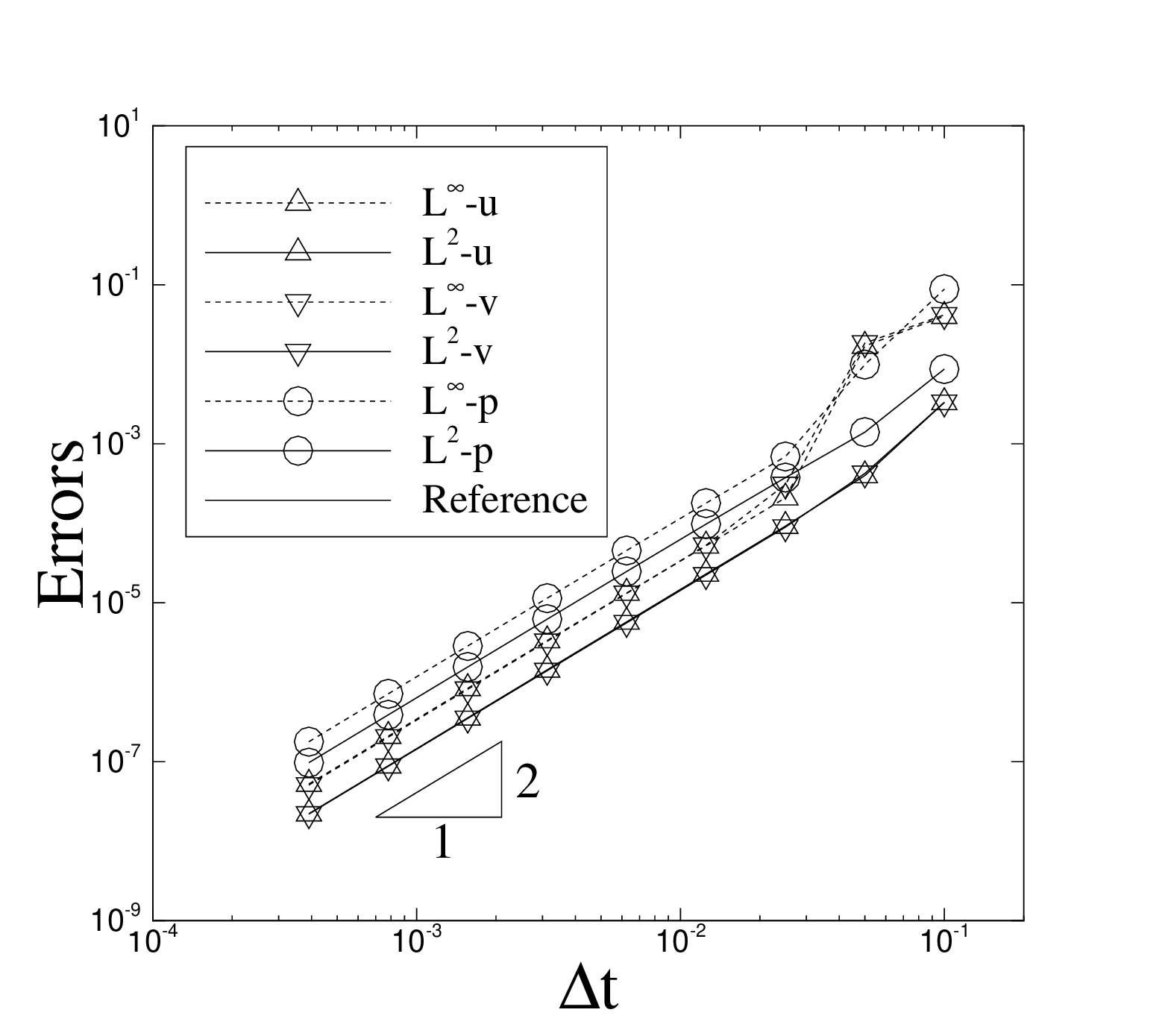}(c)
}
\caption{
Spatial/temporal convergence rates:
Flow configuration and boundary conditions (a), and
$L^\infty$ and $L^2$ errors as a function of
the element order with fixed $\Delta t=0.001$ (b)
and as a function of $\Delta t$ with a fixed element order
$16$ (c).
}
\label{fig:conv}
\end{figure}

In this subsection we study 
the spatial and temporal
convergence rates of the algorithm presented in
Section \ref{sec:algorithm} by considering an analytic
solution to the Navier-Stokes equation together
with the open boundary condition proposed in
Section \ref{sec:obc}.

Figure \ref{fig:conv}(a)
shows the problem setting.
Consider the rectangular domain
$\overline{ABDC}$ defined by
$0\leqslant x\leqslant 2$ and
$-1\leqslant y \leqslant 1$,
and the following analytic solution 
to the Navier-Stokes equations,
\eqref{equ:nse} and \eqref{equ:continuity},
\begin{equation}
\left\{
\begin{split}
&
u = 2 \cos \pi y\sin \pi x sin t \\
&
v = -2\sin\pi y \cos \pi x \sin t \\
& 
p = 2 \sin \pi y \sin \pi x \cos t
\end{split}
\right.
\label{equ:anal_soln}
\end{equation}
where $\mathbf{u}=(u,v)$.
We use a characteristic velocity
scale $U_0=1$ and a non-dimensional
viscosity $\nu=0.01$ for this problem.
The external body force $\mathbf{f}(\mathbf{x},t)$
in \eqref{equ:nse} is chosen such that
the expressions given by \eqref{equ:anal_soln}
satisfy the equation \eqref{equ:nse}.
It is noted that the analytical solution \eqref{equ:anal_soln}
employed here 
has been used for the convergence
tests in  previous 
works \cite{DongKC2014,DongS2015}.

% boundary conditions

To simulate the problem we discretize
the domain using two equal-sized quadrilateral elements
($\overline{ABFE}$ and $\overline{EFDC}$) along the $x$ direction.
On the sides $\overline{BD}$, $\overline{AB}$
and $\overline{AE}$
we impose
the Dirichlet condition \eqref{equ:dbc},
where the boundary velocity $\mathbf{w}(\mathbf{x},t)$
is chosen according to the analytic expressions
given in \eqref{equ:anal_soln}.
On the sides $\overline{EC}$ and $\overline{CD}$
we impose the open boundary condition
\eqref{equ:obc_D_3}, in which 
we set $D_0=1.0$ and $\delta=\frac{1}{20}$ and
$\mathbf{f}_b(\mathbf{x},t)$ is chosen such that
the velocity and pressure expressions given by
\eqref{equ:anal_soln} satisfy the condition 
\eqref{equ:obc_D_3} at these boundaries.

We integrate the Navier-Stokes equations 
\eqref{equ:nse}--\eqref{equ:continuity} using 
the scheme presented in \ref{sec:algorithm} in
time from $t=0$ to $t=t_f$ ($t_f$ to be specified
below). Then we compute the errors
of the numerical solution at $t=t_f$ against
the analytic expression given in \eqref{equ:anal_soln}.
The element order or the time step size
$\Delta t$ has been varied systematically,
and the errors are collected and monitored
to study the convergence behavior of
the method.

Let us first look into the spatial convergence behavior.
In this group of tests we fix the time step size at $\Delta t=0.001$ and
the integration time at $t_f=0.1$ ($100$ time steps),
and then vary the element order systematically
between $2$ and $20$. 
The numerical errors corresponding to each element order
have been computed and monitored.
Figure \ref{fig:conv}(b) shows the $L^\infty$ and $L^2$
errors of the velocity and the pressure 
as a function of the element order from these tests.
As the element order increases but within order $12$,
all the numerical errors are observed to decrease
exponentially. 
When the element order increases to $12$ and beyond,
the error curves are observed to level off
at a level $\sim 10^{-7}$ for this problem. 
The saturation of the total numerical error is because
the temporal truncation error becomes dominant when
the element order becomes large.
These results demonstrate the spatial exponential
convergence rate of our method.

% temporal convergence

The temporal convergence behavior of the method
is demonstrated by Figure \ref{fig:conv}(c), 
in which we plot the $L^\infty$ and $L^2$
errors of the flow variables as a function
of the time step size $\Delta t$.
In this group of tests the integration time
is fixed at $t_f=0.5$, the element order is fixed
at $16$, and 
$\Delta t$ is varied systematically
between $\Delta t=0.1$ and 
$\Delta t = 0.000390625$.
The convergence appears somewhat not very regular
when $\Delta t$ is above $0.025$, especially in terms of
the $L^\infty$ error norms. 
As $\Delta t$ decreases below $0.025$,
one can observe a second-order
convergence rate in time for all the flow
variables.

The results of this section suggest that for problems involving
open boundaries the method presented in 
Section \ref{sec:method} exhibits an
exponential convergence rate in space and
a second-order convergence rate in time.

% what else to discuss here?

\subsection{Flow Past a Circular Cylinder}
\label{sec:cylinder}

% vorticity distribution at Re=30, 200

\begin{figure}
\centerline{
\includegraphics[width=2in]{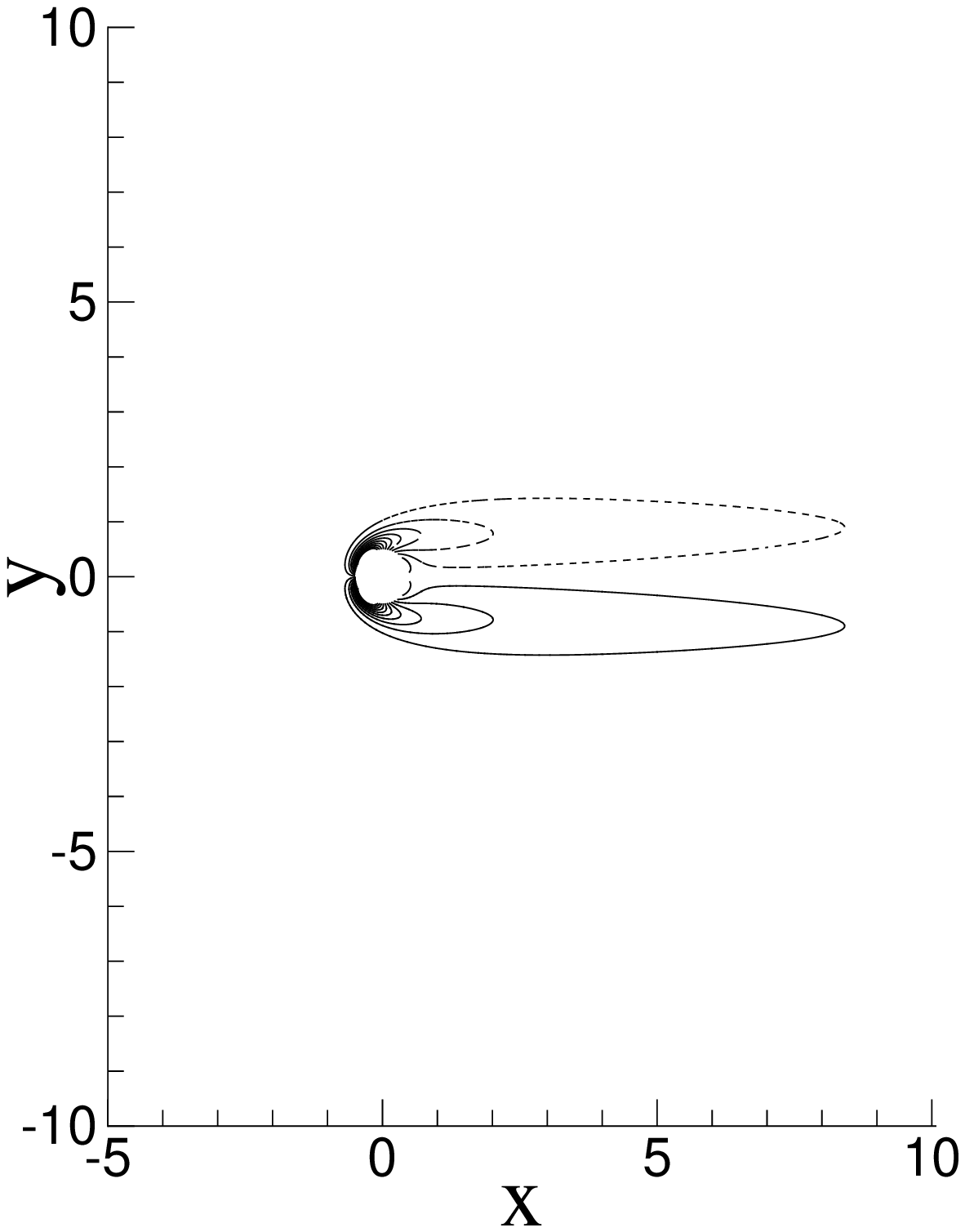}(a)
\includegraphics[width=2in]{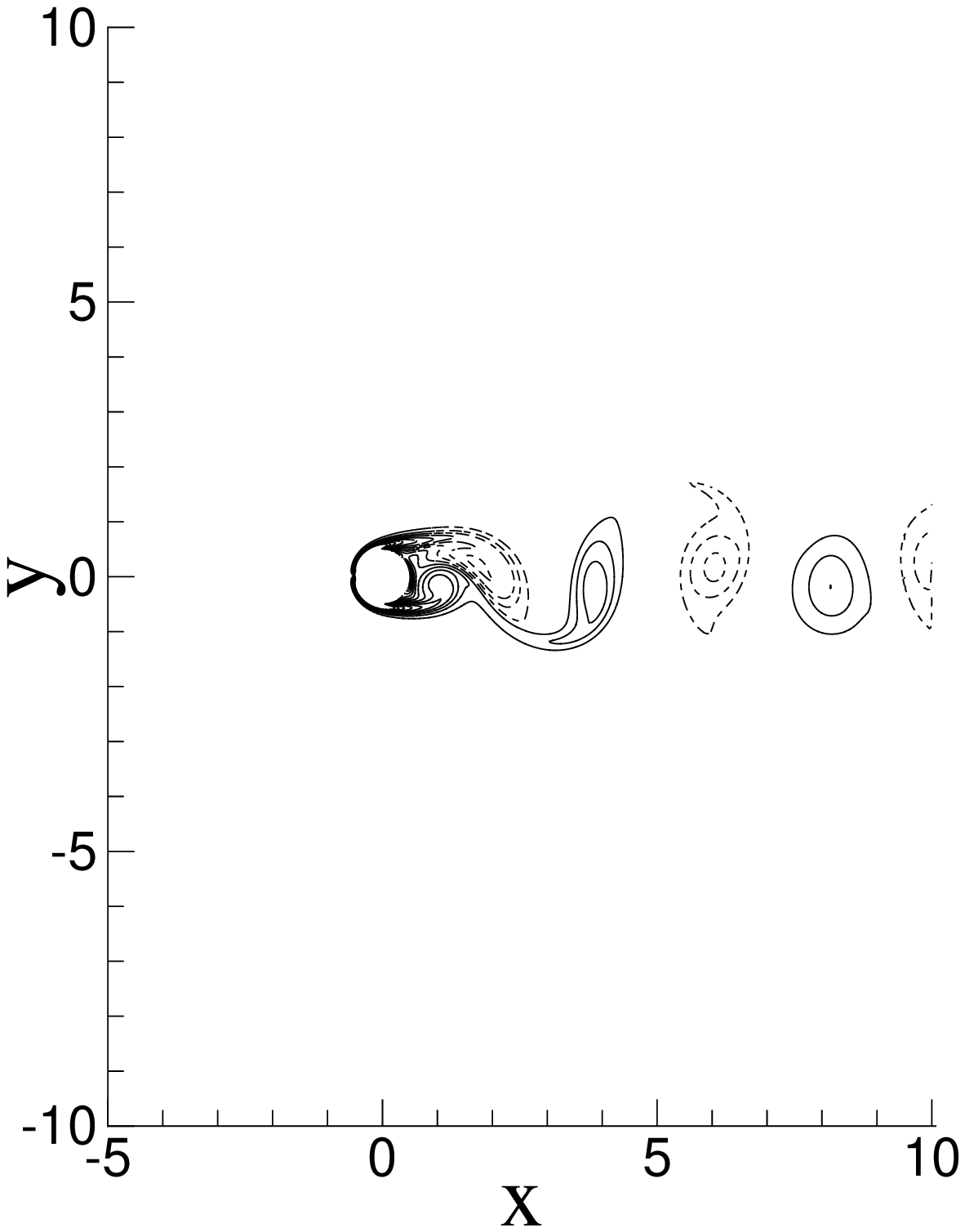}(b)
}
\caption{
%Cylinder flow: spectral element mesh (a), and
Contours of vorticity at
Reynolds numbers $Re=30$ (a) and
$Re=200$ (b). Dashed curves  indicate
negative vorticity values.
}
\label{fig:cyl_mesh_vort}
\end{figure}

In this subsection we consider a canonical wake problem, 
the flow past a circular
cylinder, in two dimensions to test the performance 
of our method.
The goal is to demonstrate the accuracy of the method
by comparison with the experimental data, and
to demonstrate its effectiveness in dealing with
the backflow instability 
as the Reynolds number becomes large.

This flow problem has been used in \cite{DongS2015}
to test a different set of open boundary 
conditions and an associated 
pressure correction-based numerical algorithm.
The flow configurations employed in the current work
largely follow those of \cite{DongS2015}.
It should be noted that the open boundary condition
and the algorithm being tested here are very
different from those of \cite{DongS2015}.

% configurations

Specifically, we consider a circular cylinder of
diameter $d$, and a
rectangular domain containing the cylinder,
$-5d\leqslant x\leqslant L$ 
and $-10d\leqslant y\leqslant 10d$,
where $x=L$ is the right domain boundary to be
specified below.
The center of the cylinder coincides with 
the origin of the coordinate system.
Four flow domains have been considered
with different wake-region sizes. 
They respectively correspond
to $L/d=5$, $10$, $15$ and $20$,
and are chosen 
in accordance with \cite{DongS2015}.
The flow domain with $L/d=10$ is illustrated
in Figure \ref{fig:cyl_mesh_vort}(a).

% boundary conditions
  
On the top and bottom domain boundaries ($y=\pm 10d$) we
assume that the flow is periodic. So
the configuration in actuality corresponds to
the flow past an infinite array of cylinders
aligned in the $y$ direction.
On the left boundary ($x=-5d$) a uniform flow
comes into the domain with
a velocity 
$\mathbf{u}=(u,v)=(U_0,0)$,
where $U_0=1$ is the characteristic
velocity scale.
The right domain boundary ($x=L$) 
is assumed to be open, where
the fluid can freely move out of the domain
and backflow may occur depending on
the flow regime and the domain size.

% how to simulate the problem?

In order to simulate the problem, we 
discretize the domain using a mesh of quadrilateral
spectral elements. 
The meshes for these four domains
respectively contain $968$, $1228$, $1488$
and $1748$ quadrilateral elements.

In simulations we impose the periodic condition
at $y/d=\pm 10$, and the  velocity Dirichlet condition
\eqref{equ:dbc}
at the inflow boundary $x=-5d$ with
a boundary velocity $\mathbf{w}=(U_0,0)$. 
%where $U_0=1$ is the free stream velocity.
On the cylinder surface a velocity no-slip condition
 is imposed, i.e.
the Dirichlet condition \eqref{equ:dbc}
with $\mathbf{w}=0$.
At the open (outflow) boundary
$x=L$ we impose the open boundary
condition \eqref{equ:obc_D_3}
with $\mathbf{f}_b=0$ and $\delta=\frac{1}{100}$.

We employ the algorithm developed
in Section \ref{sec:method} to solve the
incompressible Navier-Stokes equations.
All the length variables are normalized
by the cylinder diameter $d$ 
and the velocity is normalized by $U_0$.
So the Reynolds number for this problem 
is defined by
\begin{equation}
Re = \frac{1}{\nu} = \frac{U_0 d}{\nu_f}
\label{equ:Re}
\end{equation}
where $\nu_f$ is the kinematic viscosity 
of the fluid.
A range of Reynolds numbers (up to $Re=10000$) 
has been considered.
We use an element order $6$ for Reynolds numbers
below $100$, and an element order $8$ for 
higher Reynolds numbers.
For selected Reynolds numbers we have also performed
simulations with even larger element orders (up to $12$),
and we observe that the difference in the results when compared
with the element order $8$ is small.
The non-dimensional time step size is $U_0\Delta t/d=10^{-3}$
for Reynolds numbers below $100$ and
$U_0\Delta t/d=2.5\times 10^{-4}$ for higher Reynolds numbers
in the simulations.

% what should be the $D_0$ value in OBC?

As discussed in Section \ref{sec:method},
the analogy between the open boundary
condition \eqref{equ:convec_like_2}
and the convective boundary condition \eqref{equ:convective_obc}
suggests that $\frac{1}{D_0}$ 
represents a convection velocity. 
For the majority of simulations in this section
we employ  the average velocity at the outflow
boundary, $U_0$, as this convection velocity and 
set $D_0 = \frac{1}{U_0}$
in the open boundary condition \eqref{equ:obc_D_3}.
This is the default $D_0$ value  
for the results reported in this section.
For several selected Reynolds numbers
we have also 
investigated the effects of $D_0$
 on the simulation results.
Results corresponding to the other $D_0$ values
will be explicitly specified.

% discussion of flow regimes

The cylinder wake can be classified into several regimes,
exhibiting a variety of flow features.
These  have been expounded 
in the review paper \cite{Williamson1996}.
For Reynolds numbers below about $Re=47$
the cylinder flow is two-dimensional and  at
a steady state. 
As the Reynolds number increases beyond this
value, the cylinder wake becomes unsteady and
is characterized by vortex sheddings.
It remains two-dimensional for Reynolds numbers
up to about $Re=180$.
As the Reynolds number increases beyond
$Re\approx 180$, the cylinder wake develops
an instability and the physical flow becomes three-dimensional.
More complicated flow features and turbulence
develop in the cylinder wake 
when the Reynolds number increases further.
In Figures \ref{fig:cyl_mesh_vort}(a) and (b) 
 we plot contours of the instantaneous
vorticity obtained on the domain
$L/d=10$ at Reynolds numbers 
$Re=30$ and $Re=200$, respectively.
The results show a steady-state flow at $Re=30$
and regular vortex sheddings at $Re=200$.

% lift history Re=60 and 1000

\begin{figure}
\centering
\includegraphics[width=4.5in]{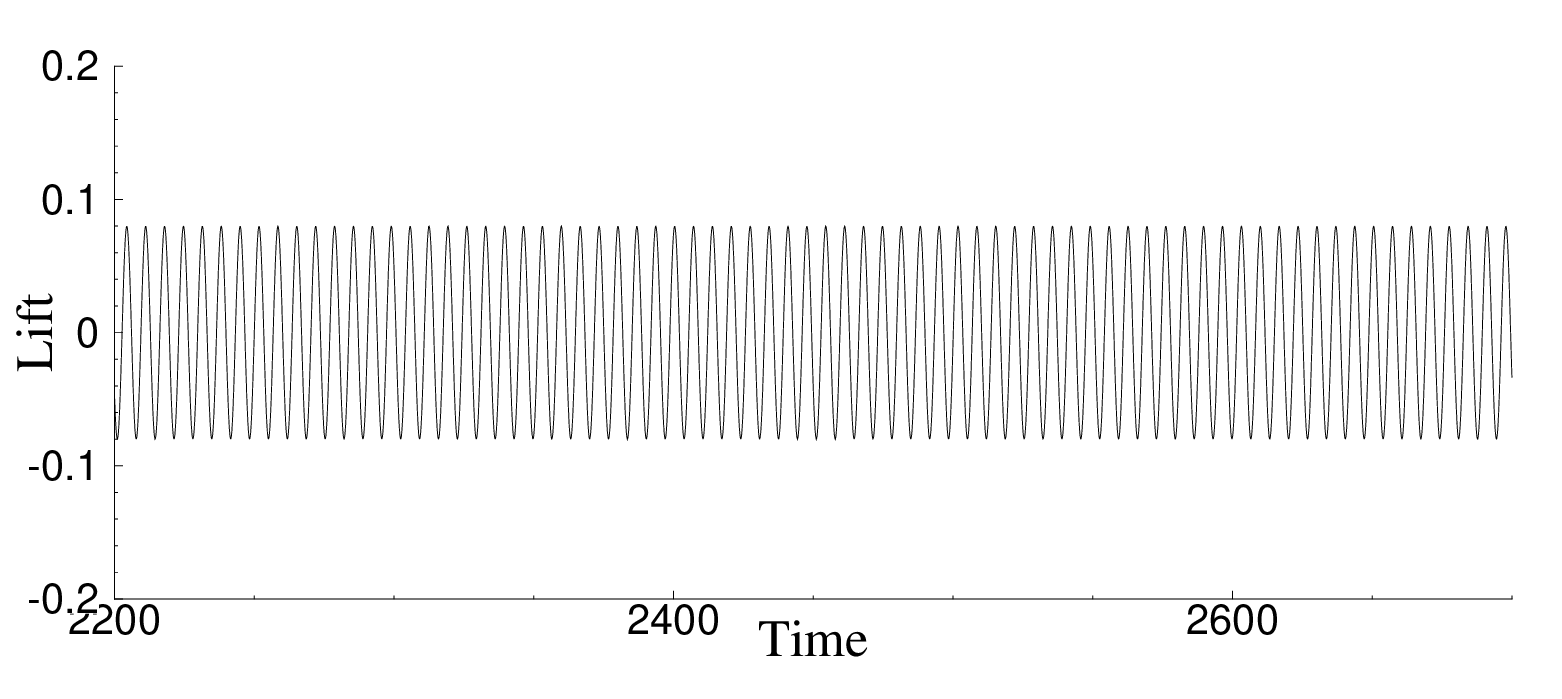}(a)
\includegraphics[width=4.5in]{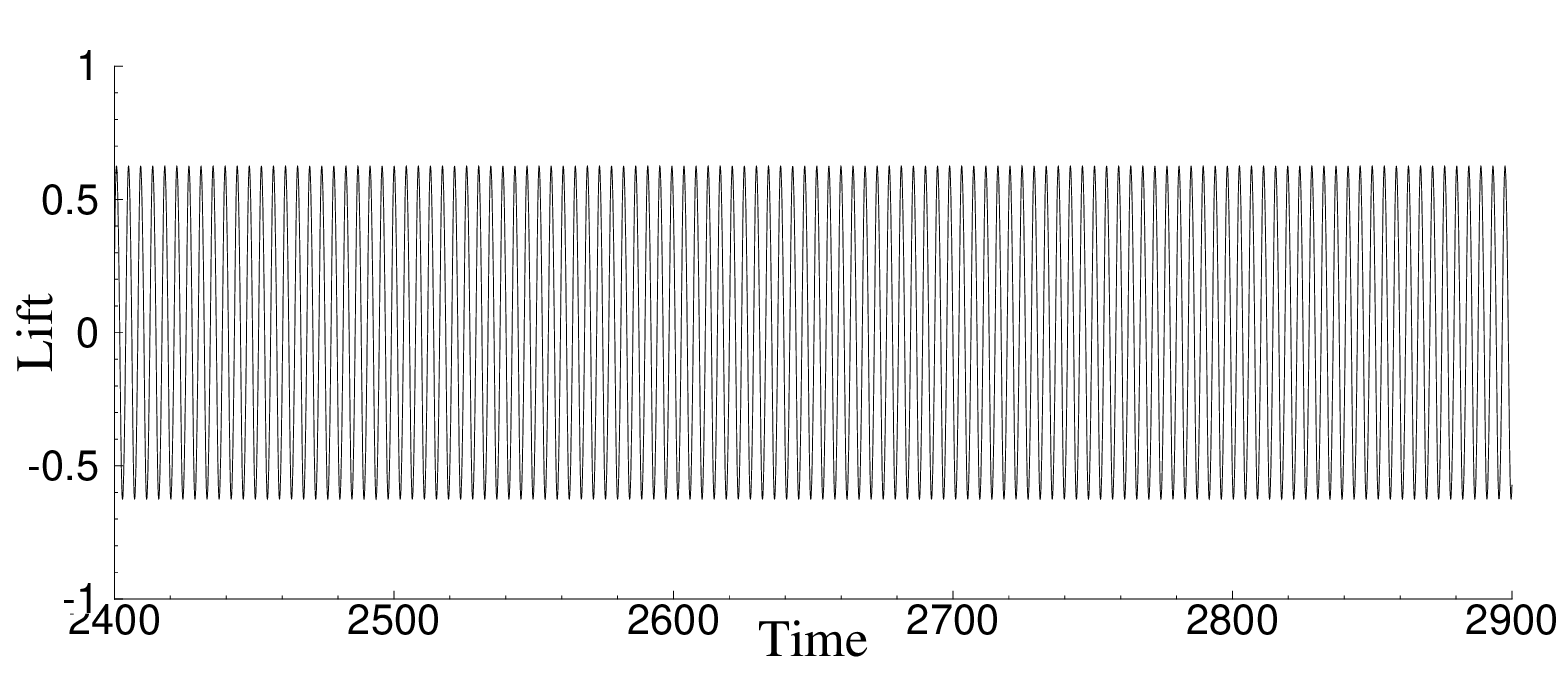}(b)
\caption{
Cylinder flow: time histories of the lift force
on the cylinder at Reynolds numbers
(a) $Re=60$ and
(b) $Re=500$.
Results are obtained with $D_0=\frac{1}{U_0}$ in the open
boundary condition.
}
\label{fig:lift_hist_low_re}
\end{figure}

We have computed and monitored the forces acting 
on the circular cylinder. 
In Figure \ref{fig:lift_hist_low_re}
we show a window of the time histories of the lift 
(i.e. the force component  in the cross-flow $y$
direction)
at Reynolds numbers 
$Re=60$ and $500$.
The force signals exhibit  quite 
regular fluctuations about a zero mean value
at these Reynolds numbers.
%

% table of force parameters at Re=20, 100
% comparing effect of wake region sizes

\begin{table}
\begin{center}
\begin{tabular*}{0.7\textwidth}{ @{\extracolsep{\fill}}  l | l | c c c  }
\hline
Reynolds number & Domain & $C_d$ & $C_d^\prime$ & $C_L$  \\ \hline
$20$ & $L/d=5$ & $2.294$ & $0$ & $0$  \\ 
     & $L/d=10$ & $2.317$ & $0$ & $0$  \\
     & $L/d=15$ & $2.317$ & $0$ & $0$  \\
     & $L/d=20$ & $2.317$ & $0$ & $0$  \\ \hline
$100$ & $L/d=5$ & $1.441$ & $8.491E-3$ & $0.261$  \\
      & $L/d=10$ & $1.459$ & $7.631E-3$ & $0.254$  \\
      & $L/d=15$ & $1.462$ & $7.700E-3$ & $0.253$  \\
      & $L/d=20$ & $1.462$ & $7.714E-3$ & $0.253$  \\ 
\hline
\end{tabular*}
\end{center}
\caption{
Cylinder flow: effect of the domain size on
the global flow parameters.
$C_d$: drag coefficient or time-averaged 
mean drag coefficient;
$C_d^\prime$: rms drag coefficient;
$C_L$: rms lift coefficient.
%$St$: Strouhal number.
}
\label{tab:force_domain_low_re}
\end{table}

Global flow parameters can be
determined based on these force data. 
In Table \ref{tab:force_domain_low_re}
we have listed several flow parameters for 
two Reynolds numbers $Re=20$ and $100$
obtained on different flow domains.
They include: the  drag coefficient
$C_d = \frac{\overline{f}_x}{\frac{1}{2}\rho U_0^2}$,
where $\overline{f}_x$ is the time averaged
drag (i.e. force component in x direction)
and $\rho=1$ is the fluid density;
the root-mean-square (rms) drag coefficient
$C_d^\prime = \frac{{f}_x^\prime}{\frac{1}{2}\rho U_0^2}$,
where $f_x^\prime$ is the rms drag;
and the rms lift coefficient 
$C_L = \frac{f_y^\prime}{\frac{1}{2}\rho U_0^2}$,
where $f_y^\prime$ is the rms lift.
These data  indicate that 
the size of the wake region has an influence on
the simulation results, and that
as the wake region becomes sufficiently large
the flow parameters computed from the simulations
remain essentially unchanged as the domain size
further increases.
It can be observed that the flow
domain $L/d=10$ is very close to 
the point where further increase in the domain
size no longer results in significant changes
in the results.
In light of this observation,
our subsequent discussions will be mainly
based on
the results obtained on the domain 
$L/d=10$.

% comparison: drag and lift coefficients

\begin{figure}
\centerline{
\includegraphics[width=3in]{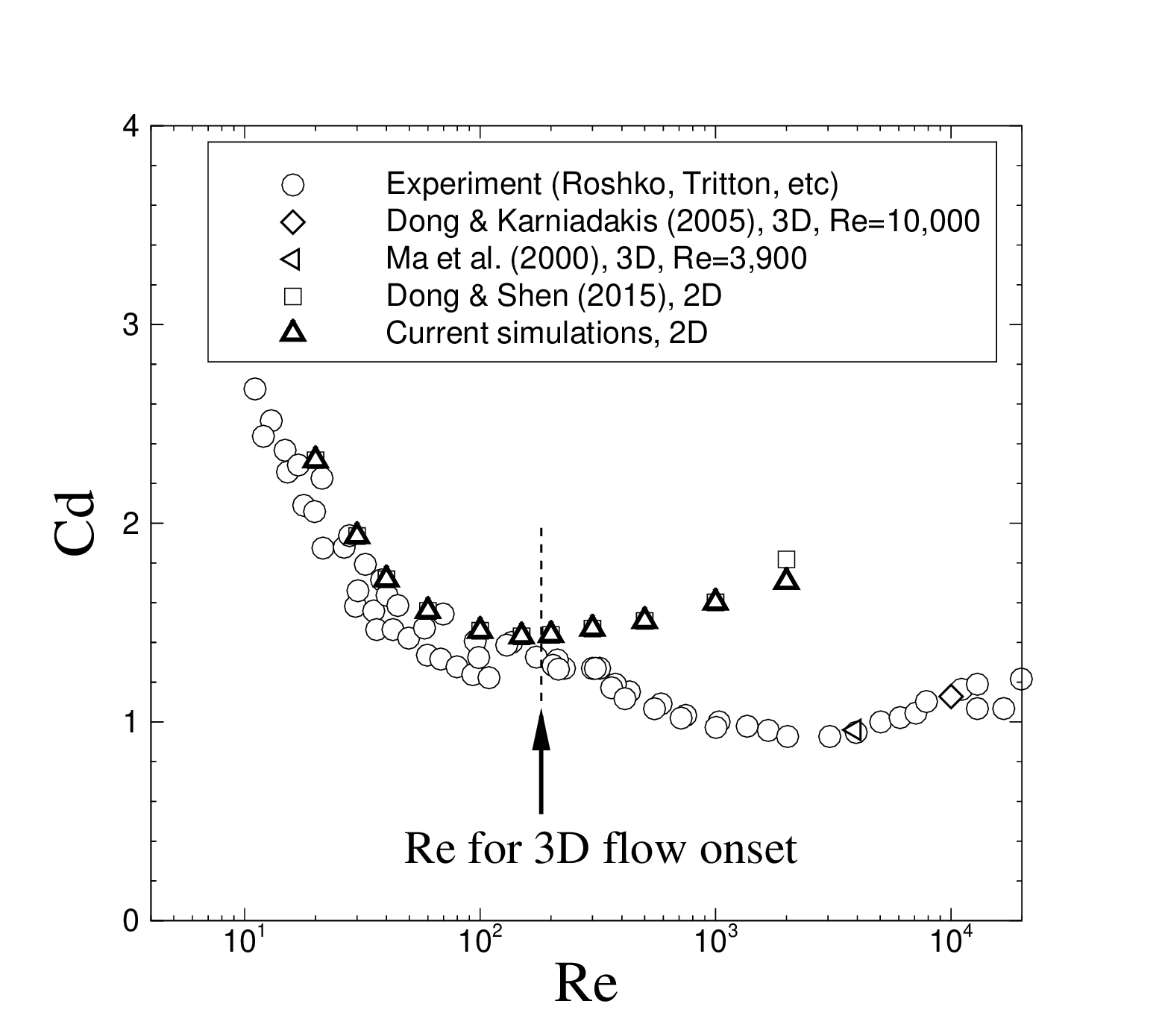}(a)
\includegraphics[width=3in]{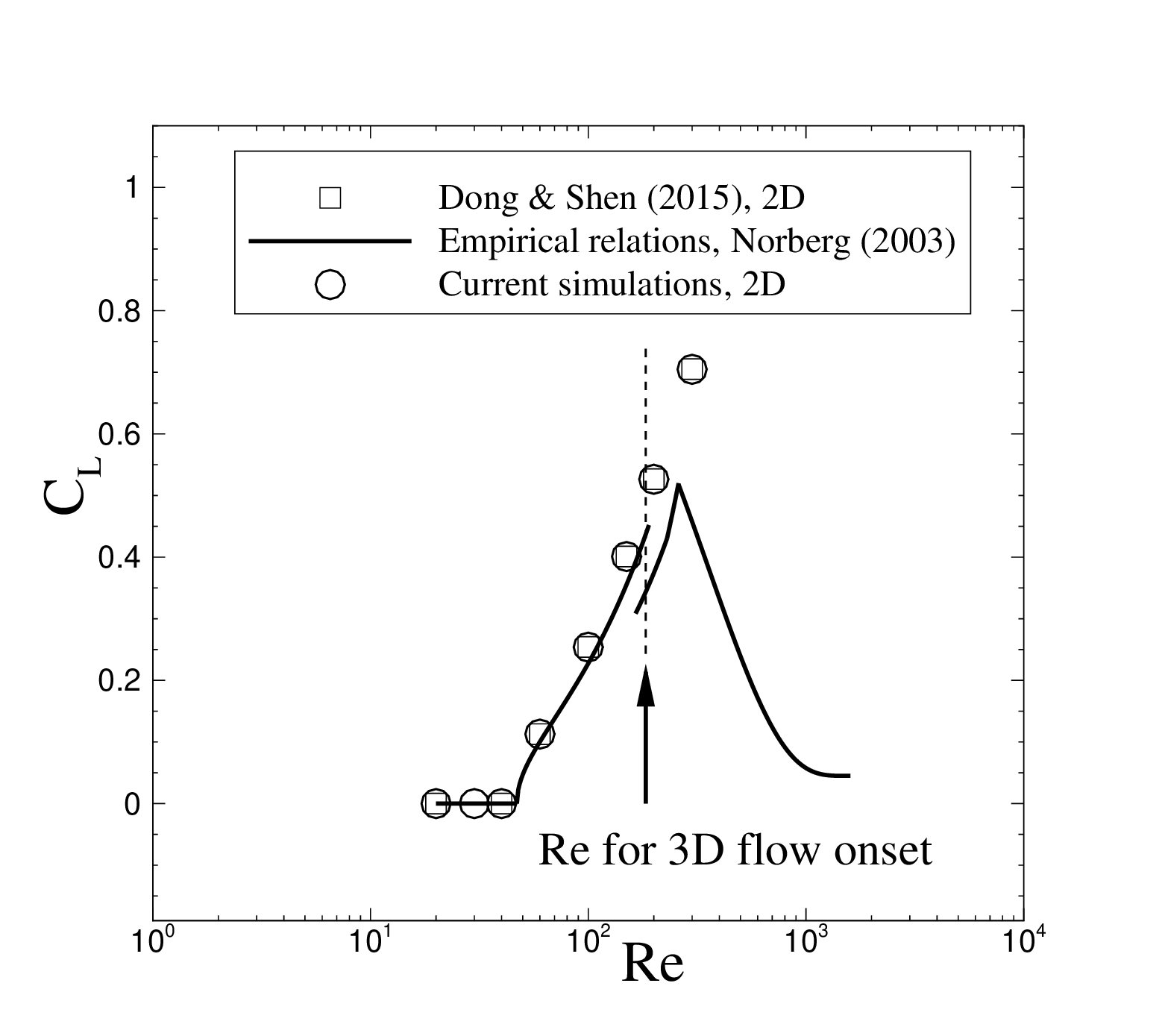}(b)
}
\caption{
Cylinder flow: Comparison of (a) drag coefficient
and (b) rms lift coefficient versus Reynolds number
between the current simulations, the experimental data,
and the simulation results of \cite{DongS2015}.
}
\label{fig:drag_lift_compare_exp}
\end{figure}

To demonstrate the accuracy of the method,
we  compare the force parameters computed
from the current simulations with those from  the experimental
measurements and from other simulations in the literature.
In Figure \ref{fig:drag_lift_compare_exp}(a)
we plot the drag coefficient ($C_d$) as a function
of Reynolds number from the
current simulations, from a number of  
experiments \cite{Wieselsberger1921,DelanyS1953,
Finn1953,Tritton1959,Roshko1961},
and from the simulations 
of \cite{MaKK2000,DongK2005,DongS2015}.
Note that the simulations in \cite{DongS2015} and
in the current work are both two-dimensional,
while those of \cite{MaKK2000,DongK2005,DongKER2006}
are three-dimensional.
The current results correspond to the
domain size $L/d=10$ and $D_0=\frac{1}{U_0}$
in the open boundary condition \eqref{equ:obc_D_3}.
They  agree with those of \cite{DongS2015}
very well. Note that both the numerical algorithm  and
the outflow boundary condition 
 in the current work are different
from those of \cite{DongS2015}.
In the two-dimensional regime
the current results also agree well
with the experimental data.
But for Reynolds numbers where the physical flow
has become three-dimensional (beyond about
$Re=180$), the current two-dimensional
simulations result in overly large drag coefficients
compared to the experiments,
and the discrepancy grows with increasing Reynolds number.

Figure \ref{fig:drag_lift_compare_exp}(b) is
a comparison of the rms lift coefficient $C_L$
as a function of the Reynolds number
between current simulations, the experiment of \cite{Norberg2003},
and the simulations of \cite{DongS2015}.
The curves show the empirical relation given by
\cite{Norberg2003} based on several experimental
sources, which exhibits a hysteresis around the
Reynolds numbers where the two-dimensional to
three-dimensional flow transition occurs.
The lift coefficients from the current simulations and from
\cite{DongS2015} agree with each other almost exactly.
In the two-dimensional regime the current results
agree with the empirical relation
from \cite{Norberg2003} reasonably well.
In the three-dimensional regime, however, 
the current two-dimensional
simulations grossly over-predict the lift coefficient,
which is a well-known phenomenon about two-dimensional
simulations (see e.g. \cite{DongK2005,DongKER2006}).

% table of lift coefficients

\begin{table}
\begin{center}
\begin{tabular*}{0.7\textwidth}{@{\extracolsep{\fill}}
l c c}
\hline
Reference & $Re=100$ & $Re=200$ \\
Braza et al. (1986) \cite{BrazaCH1986} & $0.21$ & $0.55$ \\
Engelman \& Jamnia (1990) \cite{EngelmanJ1990} & $0.26$ & -- \\
Meneghini \& Bearman (1993) \cite{MeneghiniB1993} & -- & $0.54$ \\
Beaudan \& Moin (1994) \cite{BeaudanM1994} & $0.24$ & -- \\
Zhang et al. (1995) \cite{Zhangetal1995} & $0.25$ & $0.53$ \\
Tang \& Audry (1997) \cite{TangA1997} & $0.21$ & $0.45$ \\
%Zhang \& Dalton (1997) & -- & $0.38$ \\
Persillon \& Braza (1998) \cite{PersillonB1998} & $0.27$ & $0.56$ \\
Zhang \& Dalton (1998) \cite{ZhangD1998} & -- & $0.48$ \\
Kravchenko et al. (1999) \cite{KravchenkoMS1999} & $0.22$ & -- \\
Hwang \& Lin (1992) \cite{HwangL1992} & $0.27$ & $0.42$ \\
Franke et al. (1990) \cite{FrankeRS1990} & -- & $0.46$ \\
Karniadakis (1988) \cite{Karniadakis1988} & -- & $0.48$ \\
Newman \& Karniadakis (1995) \cite{NewmanK1995} & -- & $0.51$ \\
Newman \& Karniadakis (1996) \cite{NewmanK1996} & $0.24$ & -- \\
Dong \& Shen (2010) \cite{DongS2010} & -- & $0.501$ \\
Dong \& Shen (2015) \cite{DongS2015}  & $0.254$ & $0.527$ \\
Current simulation (domain $L/d=10$) & $0.254$ & $0.526$ \\
Current simulation (domain $L/d=20$) & $0.253$ & -- \\
\hline
\end{tabular*}
\caption{ Cylinder flow: Comparison of
rms lift coefficients at $Re=100$ and $Re=200$
between current simulations and other  simulations
from literature.
}
\label{tab:circyl_lift_re100}
\end{center}
\end{table}

In Table \ref{tab:circyl_lift_re100} we have 
summarized the rms lift coefficients ($C_L$)
for Reynolds numbers $Re=100$ and $200$
from a number of two-dimensional simulations
in the literature. We have also listed
the $C_L$ values on flow domains with $L/d=10$ and $20$
from the current simulations for comparison;
see Table \ref{tab:force_domain_low_re}
for $C_L$ values on the other domains at $Re=100$.
One can observe a spread in the $C_L$ values
from different simulations.
The lift coefficients from the current work
are well within the range of values 
from the literature.

% what else to discuss in terms of accuracy?

% velocity field
\begin{figure}
\centering
\includegraphics[width=4.in]{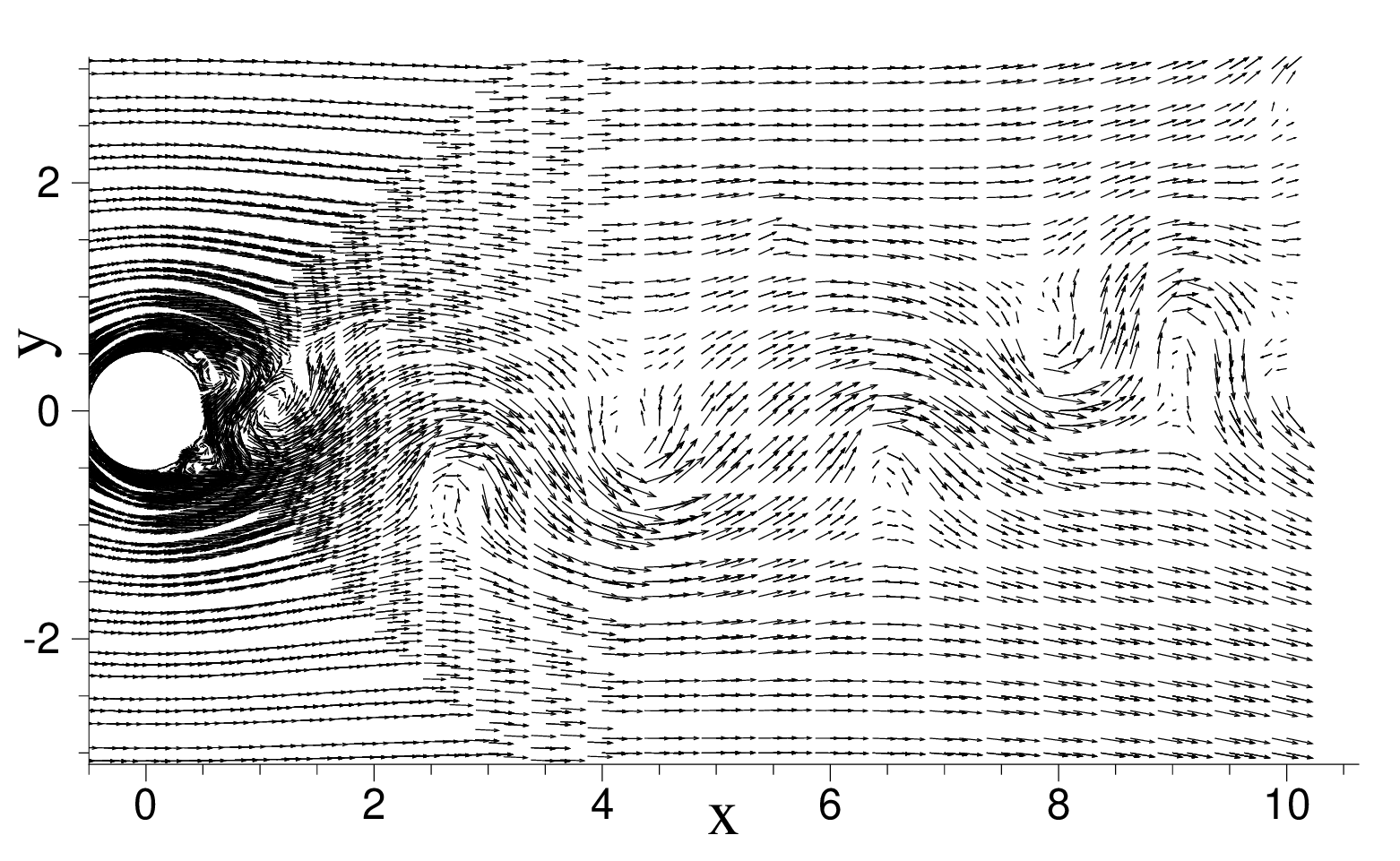}(a)
\includegraphics[width=4.in]{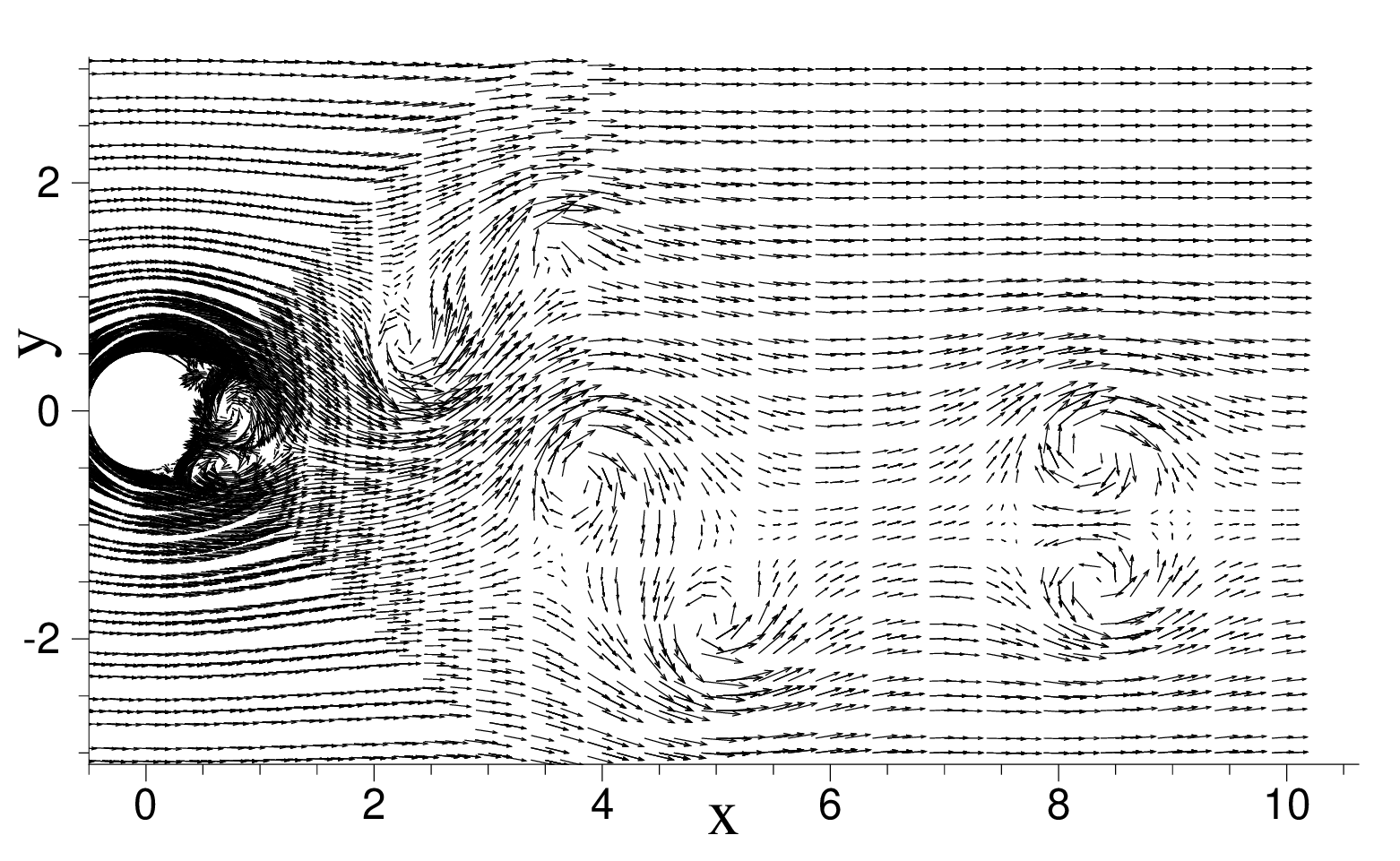}(b)
\includegraphics[width=4.in]{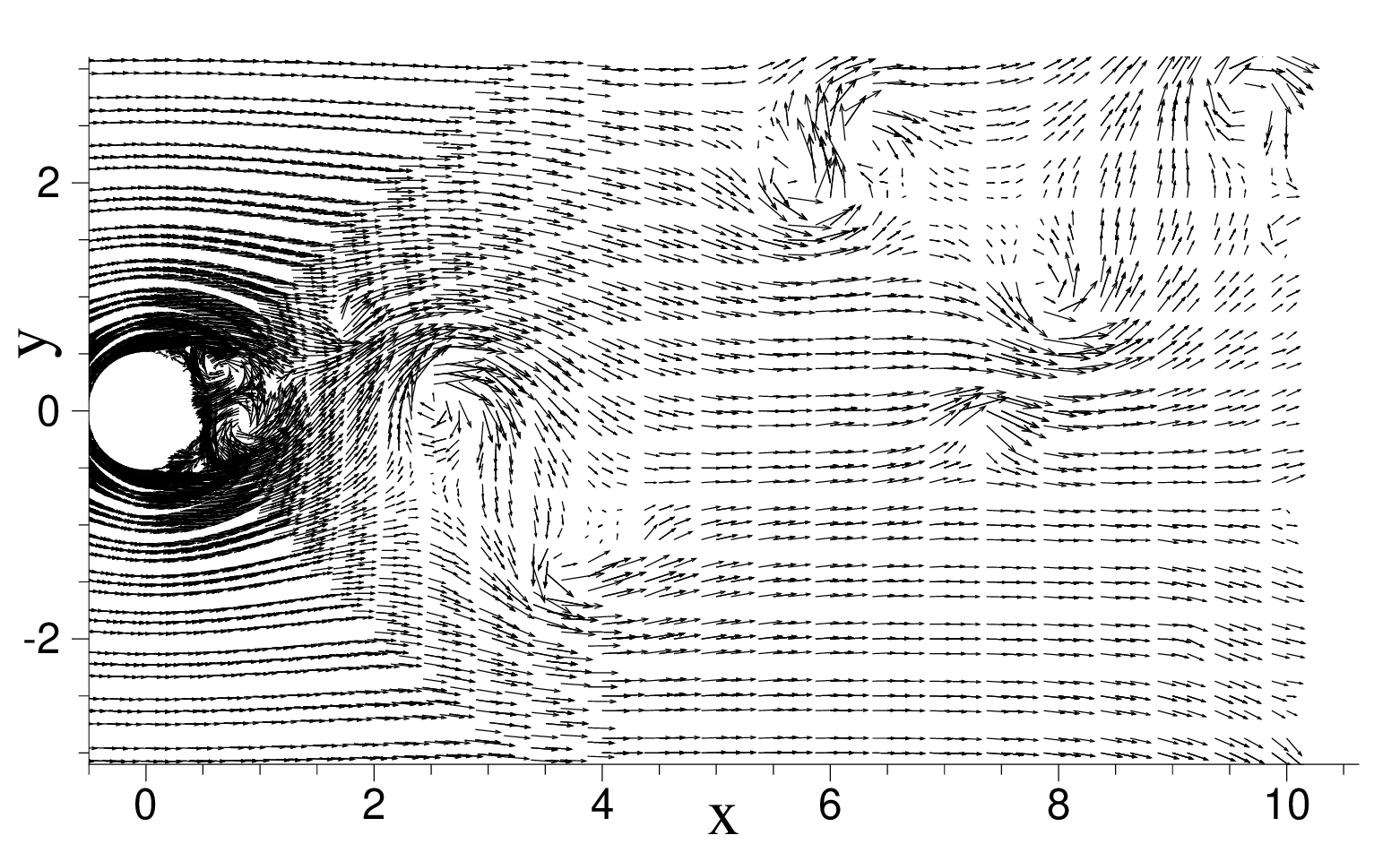}(c)
\caption{
Cylinder flow:
snapshots of instantaneous velocity fields at
Reynolds numbers 
(a) $Re=2000$,
(b) $Re=5000$, and
(c) $Re=10000$.
Velocity vectors are plotted on every fifth quadrature
points in each direction within each element.
}
\label{fig:cyl_vel}
\end{figure}

% stability and effectiveness of OBC for higher Re

Let us next look into the effectiveness of
the open boundary condition and the algorithm from Section
\ref{sec:method} for dealing with
the backflow instability.
We will consider the cylinder flow at
higher Reynolds numbers, ranging from
$Re=2000$ to $Re=10000$.
At these Reynolds numbers 
energetic vortices are observed to pass through 
the outflow boundary and induce strong backflows 
in that region.
This creates a severe instability issue, and
makes the simulation immensely challenging.

Thanks to the energy stability, the current open boundary
condition provides an effective way for
overcoming this instability.
In Figure \ref{fig:cyl_vel} 
we  show distributions of the
instantaneous velocity at three
Reynolds numbers $Re=2000$, $5000$ 
and $10000$.
The results are obtained on the 
domain $L/d=10$, with
$D_0=\frac{1}{U_0}$ in the open boundary
condition \eqref{equ:obc_D_3}.
Energetic vortices can be
clearly observed at the open
boundary (see e.g. Figure \ref{fig:cyl_vel}(c)).

%% lift time histories

\begin{figure}
\centering
\includegraphics[width=4.5in]{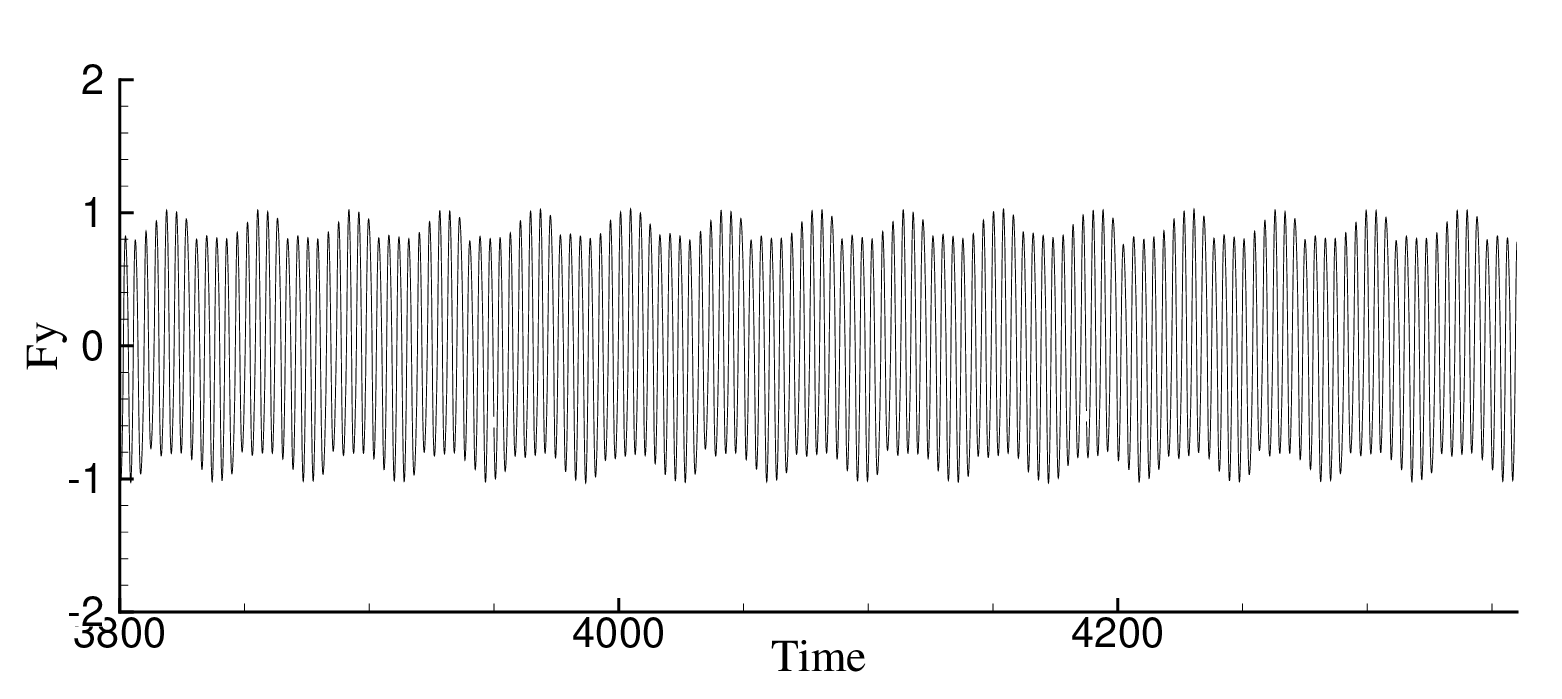}(a)
\includegraphics[width=4.5in]{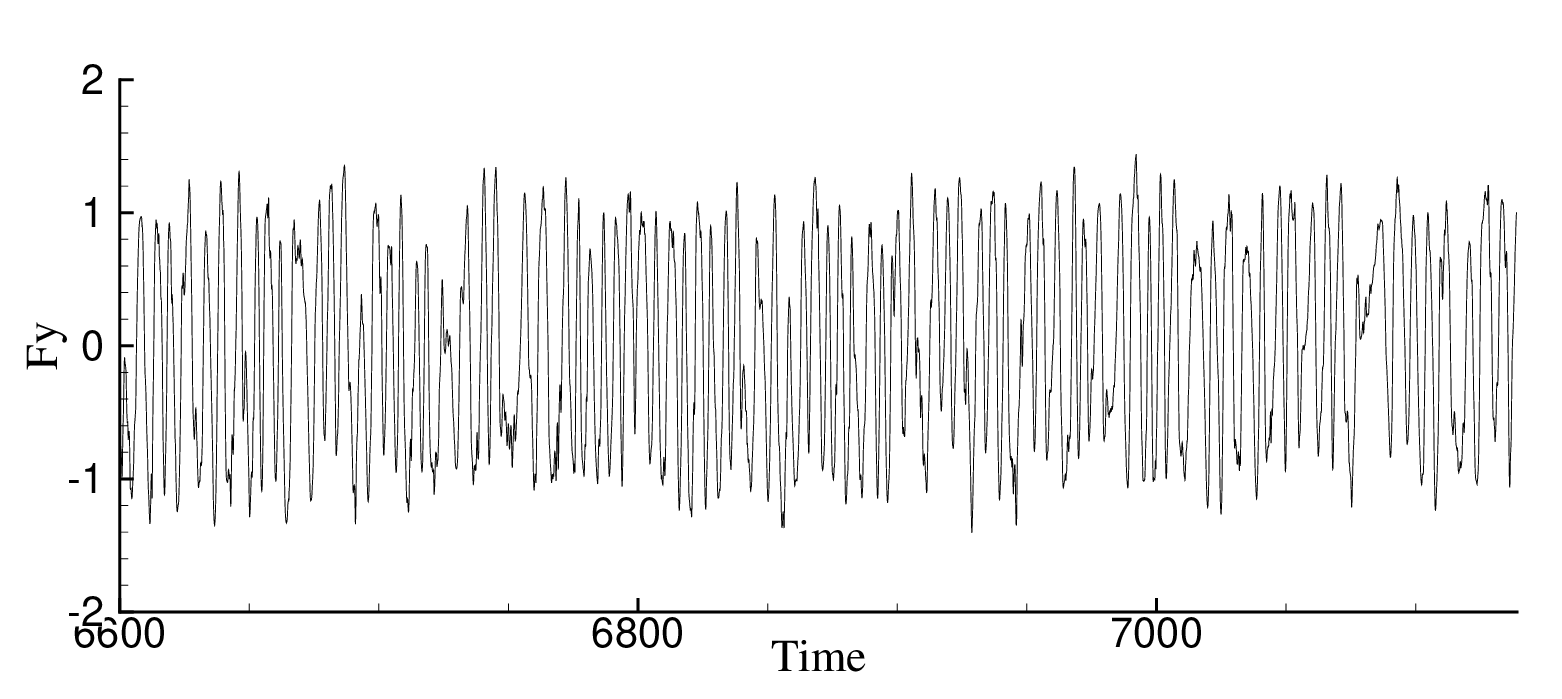}(b)
\caption{
Time histories of the lift force on the cylinder at
Reynolds numbers
(a) $Re=2000$, and
(b) $Re=10000$.
}
\label{fig:cyl_force_hist_high_re}
\end{figure}

We have performed long-time simulations at these Reynolds numbers
using the current method.
The long-term stability of the simulations
is demonstrated by Figure \ref{fig:cyl_force_hist_high_re},
in which we show a window of the time histories 
(over $30$ flow-through time)
of the lift force on the cylinder at
Reynolds numbers $Re=2000$ and $Re=10000$
obtained on the domain $L/d=10$.
At $Re=2000$ the lift signal exhibits a modulation
in its amplitude. 
At $Re=10000$ the fluctuation becomes quite chaotic
and the vortex-shedding frequency appears to vary 
over time.
These results show that simulations using
the current method are long-term stable 
in the presence of strong vortices and backflows
at the outflow/open boundaries.

In contrast, the boundary condition
\eqref{equ:convec_like_1} is observed to be unstable
at these Reynolds numbers considered here.
The computation instantly blows up as the vortices
hit the outflow/open boundary.

% drag histories with different D0

\begin{figure}
\centering
\includegraphics[width=4.5in]{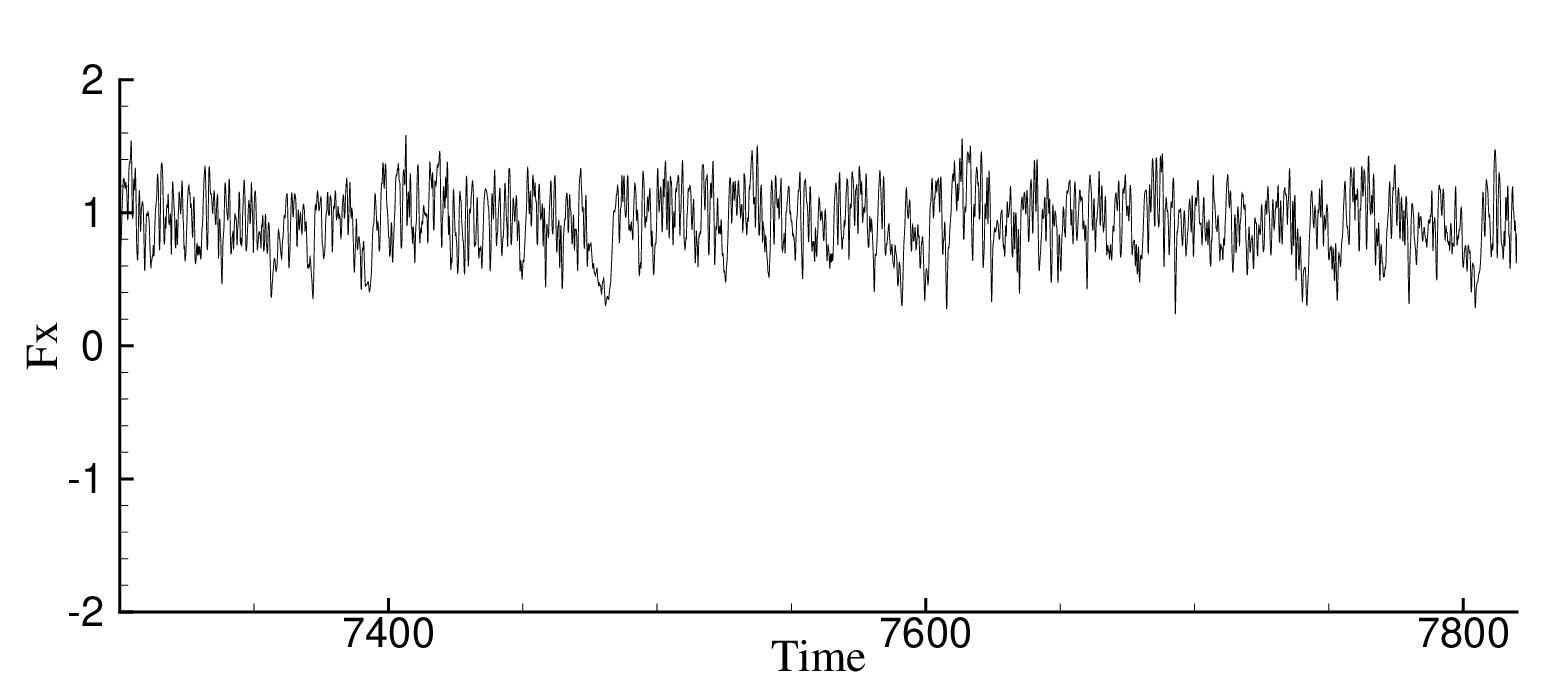}(a)
\includegraphics[width=4.5in]{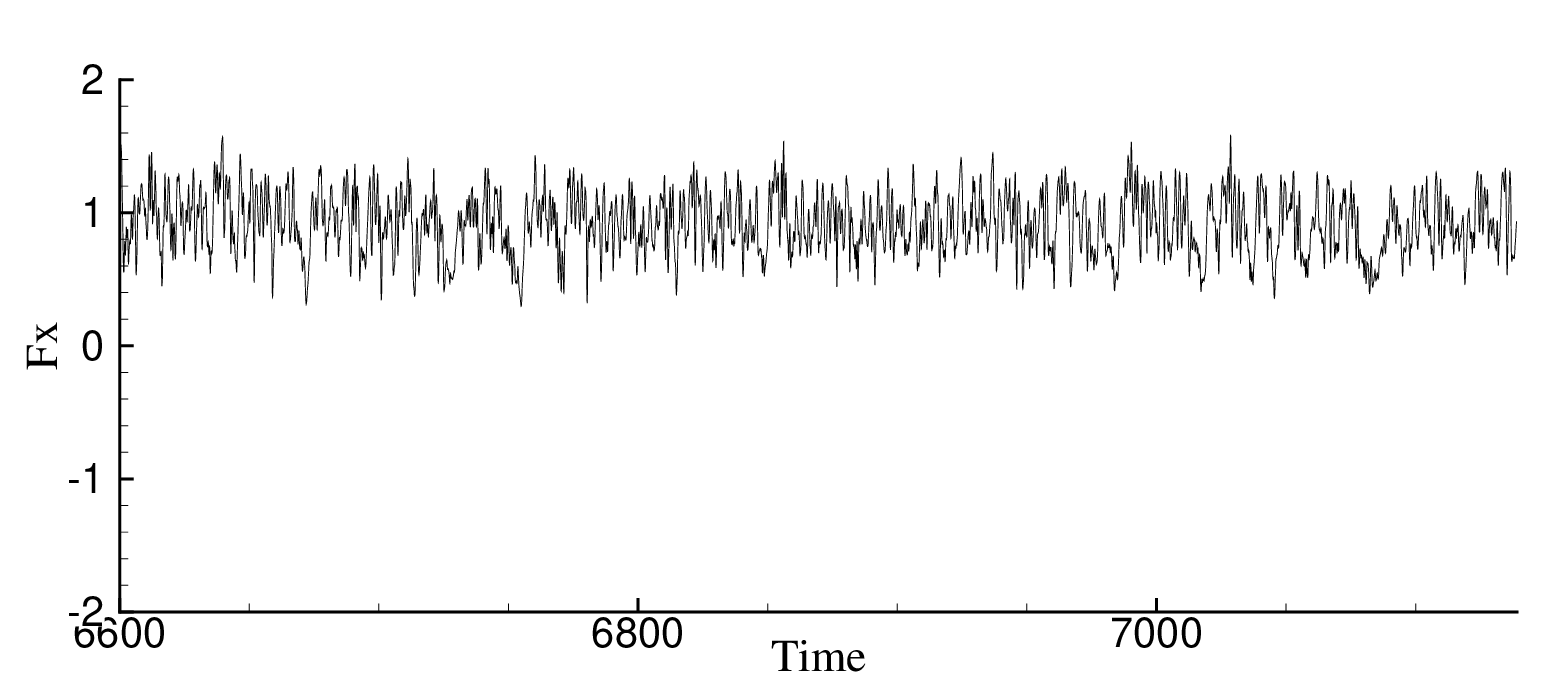}(b)
\caption{
Time histories of the drag on the cylinder at $Re=10000$
obtained using different $D_0$ values in OBC:
(a) $D_0=0$, and
(b) $D_0=\frac{1}{U_0}$.
}
\label{fig:drag_hist_D0_re10k}
\end{figure}

% effects of D_0 value in OBC
% what is the issue?
% how are you going to approach it?
% what is the result?

Let us next look into the effects of 
the $D_0$ value in the open boundary
condition \eqref{equ:obc_D_3} on
the simulation results.
By an analogy between 
the current open boundary condition and the
usual convective boundary condition
\eqref{equ:convective_obc},
we observe that
$\frac{1}{D_0}$ should correspond 
to a convection velocity scale $U_c$
at the outflow boundary, i.e.
$\frac{1}{D_0}=U_c$,
as is discussed in Section \ref{sec:method}.
The simulation results for the
cylinder flow presented
 so far are
obtained with a value
$\frac{1}{D_0}=U_c=U_0$,
where $U_0$ is the average velocity at 
the outflow boundary.

% table of global parameters for different D0 values

\begin{table}[h]
\begin{center}
\begin{tabular*}{0.7\textwidth}{@{\extracolsep{\fill}} l | l | c c c  }
\hline
Reynolds number & $D_0U_0$ & $C_d$ & $C_d^\prime$ & $C_L$ \\ \hline
$20$ & $0.0$ & $2.317$ & $0.0$ & $0.0$ \\
     & $0.5$ & $2.317$ & $0.0$ & $0.0$ \\
     & $1.0$ & $2.317$ & $0.0$ & $0.0$ \\
     & $2.0$ & $2.317$ & $0.0$ & $0.0$ \\
     & $5.0$ & $2.317$ & $0.0$ & $0.0$ \\
\hline
$100$ & $0.0$ & $1.459$ & $7.627E-3$ & $0.254$ \\
      & $0.5$ & $1.459$ & $7.630E-3$ & $0.254$ \\
      & $1.0$ & $1.459$ & $7.631E-3$ & $0.254$ \\
      & $2.0$ & $1.459$ & $7.639E-3$ & $0.254$ \\
      & $5.0$ & $1.459$ & $7.656E-3$ & $0.254$ \\
\hline
$10000$ & $0.0$ & $1.862$ & $0.449$ & $1.483$ \\
        & $0.5$ & $1.881$ & $0.421$ & $1.506$ \\
        & $1.0$ & $1.843$ & $0.442$ & $1.474$ \\
        & $2.0$ & $1.893$ & $0.432$ & $1.504$ \\
        & $5.0$ & $1.858$ & $0.446$ & $1.474$ \\
\hline
\end{tabular*}
\end{center}
\caption{
Effect of $D_0$ in OBC on global flow parameters:
drag and lift coefficients for cylinder flow 
obtained with 
different $D_0$ values. 
$C_d$: drag coefficient or time-averaged mean drag coefficient;
$C_d^\prime$: root-mean-square (rms) drag coefficient;
$C_L$: rms lift coefficient.
}
\label{tab:cyl_D0}
\end{table}

We have observed that the 
variation in the $D_0$ value in
the open boundary condition 
\eqref{equ:obc_D_3} has little
or essentially no effect 
on the global physical quantities
such as the forces on
the cylinder.
This is illustrated in
Figure \ref{fig:drag_hist_D0_re10k}
with the time histories of the
drag at Reynolds number $Re=10000$
corresponding to two
values $D_0U_0=0$ (or $U_c=\infty$) and $D_0U_0=1$ (or $U_c=U_0$)
in the open boundary condition
\eqref{equ:obc_D_3}.
The two drag signals exhibit qualitatively similar
characteristics.
A quantitative comparison of 
the global physical quantities
corresponding to several $D_0$ values
is given in Table \ref{tab:cyl_D0}.
The table includes the data
for the mean drag coefficient ($C_d$),
rms drag coefficient ($C_d^\prime$),
and the rms lift coefficient ($C_L$)
for Reynolds numbers 
$Re=20$, $100$ and $10000$
on the flow domain with $L=10d$.
We have considered several $D_0$
values in these tests, corresponding to 
$D_0U_0=0$, $0.5$, $1$, $2$ and $5$,
or equivalently 
$U_c=\infty$, $2U_0$, $U_0$,
$\frac{U_0}{2}$ and $\frac{U_0}{5}$.
%
% what is the result? what does it mean?
One can observe that 
at $Re=20$ and $Re=100$
these global physical quantities
obtained using these several $D_0$ values
are exactly or almost exactly
the same.
At $Re=10000$, they are also close
for different $D_0$ values,
with a maximum difference of about $2.7\%$
for $C_d$, about $6.7\%$ for $C_d^\prime$,
and about $2.1\%$ for $C_L$.
These results suggest that
the value of $D_0$ in the
open boundary condition \eqref{equ:obc_D_3} 
has little effect quantitatively
on the physical quantities of the flow.

The main effect of $D_0$ appears to be
on the qualitative features of the flow, such
as the smoothness of the velocity field,
in regions local to the outflow boundary.
We observe that the current open
boundary condition \eqref{equ:obc_D_3}
with appropriate $D_0>0$  
 result in smoother velocity
distributions at the outflow boundary and 
can allow the vortices to cross the 
outflow boundary more smoothly and in a more natural
way when compared to the boundary condition with $D_0=0$,
which corresponds exactly to the boundary
condition OBC-C studied in \cite{DongS2015}.
These observations suggest that 
one should choose $\frac{1}{D_0}$
in accordance with the convection velocity scale
of the vortices at the outflow boundary,
which for the current problem is close to $U_0$, 
in order to improve the qualitative characteristics
of the computed flow 
near the outflow boundary.
A larger deviation of the chosen $\frac{1}{D_0}$ value
in the open boundary condition
from the actual vortex-convection
velocity  at the outflow boundary
may lead to qualitatively poorer flow patterns
near the outflow boundary. 

% how to understand the above observations
% based on energy stability relation?
% D_0 != 0 provides control over the
%   the velocity at the outflow boundary

% what else to discuss here about cylinder flow?

\subsection{Jet in  Open Domain}
\label{sec:jet}

\begin{figure}
\centerline{
\includegraphics[height=3.in]{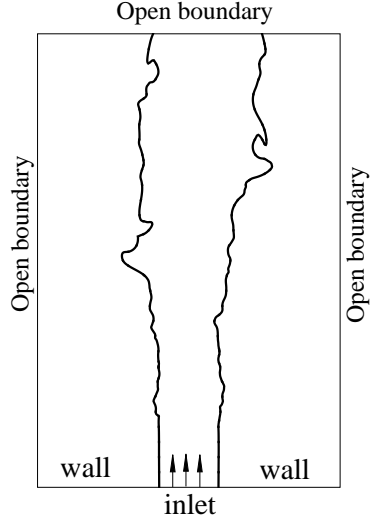}
}
\caption{
Configuration of jet in an open domain.
}
\label{fig:jet_config}
\end{figure}

In this subsection we apply the current method 
to simulate a jet in an open domain.
This is a type of flow different than that of
the previous subsection.
The open domain boundaries combined with
the physical instability of the jet at
large Reynolds numbers make this type of flows
very challenging to simulate.
We will again consider two-dimensional
simulations. 

Figure \ref{fig:jet_config} is
a sketch of the flow configuration.
We consider a jet with an inlet diameter $d$,
which is contained in a rectangular domain,
$-2.5d\leqslant x\leqslant 2.5d$
and 
$0\leqslant y\leqslant 7.5d$.
The bottom of the domain ($y=0$)
is a solid wall, and the jet inlet
is located in the middle of the wall,
covering $-0.5d\leqslant x\leqslant 0.5d$.
The other three boundaries (top, left and right) of 
the domain are open, where the fluid
can freely leave or enter the domain.
We assume that the jet velocity at
the inlet has the following profile
\begin{equation}
\left\{
\begin{split}
& u = 0 \\
&
v = U_0\left[
 \left(
   H(x,0) - H(x,d/2)
 \right) 
 \tanh \frac{1-\frac{2x}{d}}{\sqrt{2}\epsilon}
 + \left(
   H(x,-d/2) - H(x,0)
 \right)
 \tanh \frac{1+\frac{2x}{d}}{\sqrt{2}\epsilon}
\right]
\end{split}
\right.
\label{equ:jet_inlet}
\end{equation}
where $(u,v)$ are components of the
velocity $\mathbf{u}$ in $x$ and $y$
directions, $U_0$ is a velocity scale,
and $\epsilon=\frac{d}{40}$.
$H(x,a)$ is the Heaviside step function,
assuming a unit value if $x\geqslant a$
and vanishing otherwise.
Note that with this profile
the  inlet flow
has a bulk velocity $U_0$.
We assume that
there is no external body force
for this problem.

We normalize all the length variables by 
the jet inlet diameter $d$, and all the
velocity variables by $U_0$.
The Reynolds number is therefore
given by equation \eqref{equ:Re},
in which $d$ and $U_0$ have meanings
particular to this problem.

% how to simulate problem?
% what are the boundary conditions?
% what are the simulation parameters?

The domain has been discretized using
$600$ equal-sized quadrilateral spectral elements,
with $20$ elements in the 
$x$ direction and $30$ elements
in the $y$ direction.
%
% BCs
We impose the velocity Dirichlet 
condition \eqref{equ:dbc} on
the bottom side of the domain,
with the boundary velocity 
$\mathbf{w}(\mathbf{x},t)=0$ in the wall region
and  set according to
equation \eqref{equ:jet_inlet}
in the jet inlet.
On the top, left and right sides of
the domain we impose the
open boundary condition \eqref{equ:obc_D_3}
with $\mathbf{f}_b=0$ and $\delta=\frac{1}{100}$.
The majority of simulations in this subsection
are performed using $D_0 =\frac{1}{U_0}$
in the open boundary condition.
Several other $D_0$ values have also 
been considered in selected cases
for comparison.

% simulation parameters
% dt, element order, number of time steps
% 

We integrate the Navier-Stokes equations 
\eqref{equ:nse}--\eqref{equ:continuity} in time
using the algorithm described in Section \ref{sec:method}.
Long-time simulations have been performed
at three Reynolds numbers: 
$Re=2000$, $5000$ and $10000$.
In the simulations we employ an element order $12$ for each
element at the two lower Reynolds numbers,
and an element order $16$ for each element
at $Re=10000$.
The non-dimensional time step size
is $\frac{U_0\Delta t}{d}=2.5\times 10^{-4}$
for $Re=2000$ and $5000$, and
$\frac{U_0\Delta t}{d}=2.0\times 10^{-4}$
for $Re=10000$.

% what else about setup to discuss here?

\begin{figure}
\centerline{
\includegraphics[width=2.in]{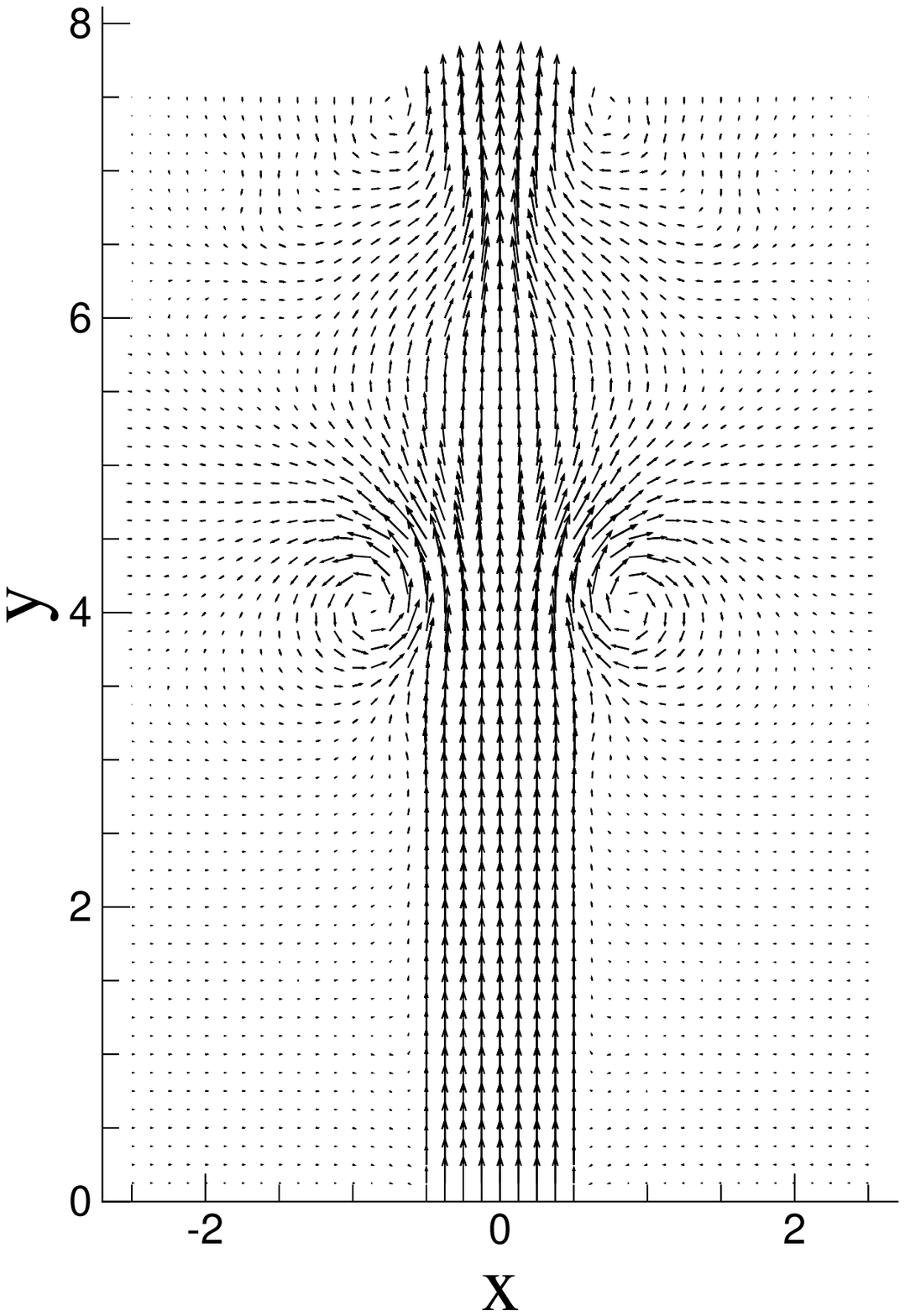}(a)
\includegraphics[width=2.in]{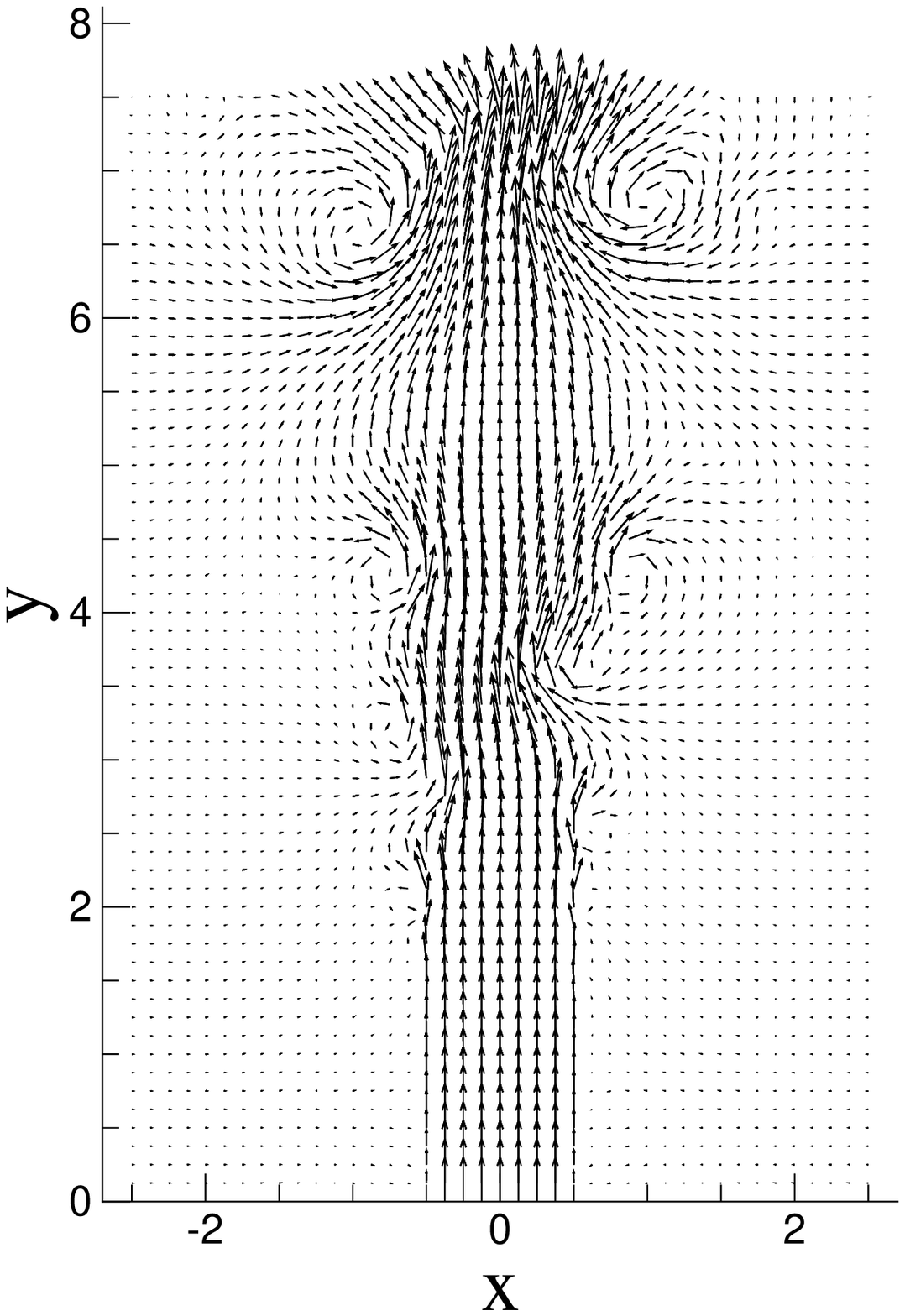}(b)
\includegraphics[width=2.in]{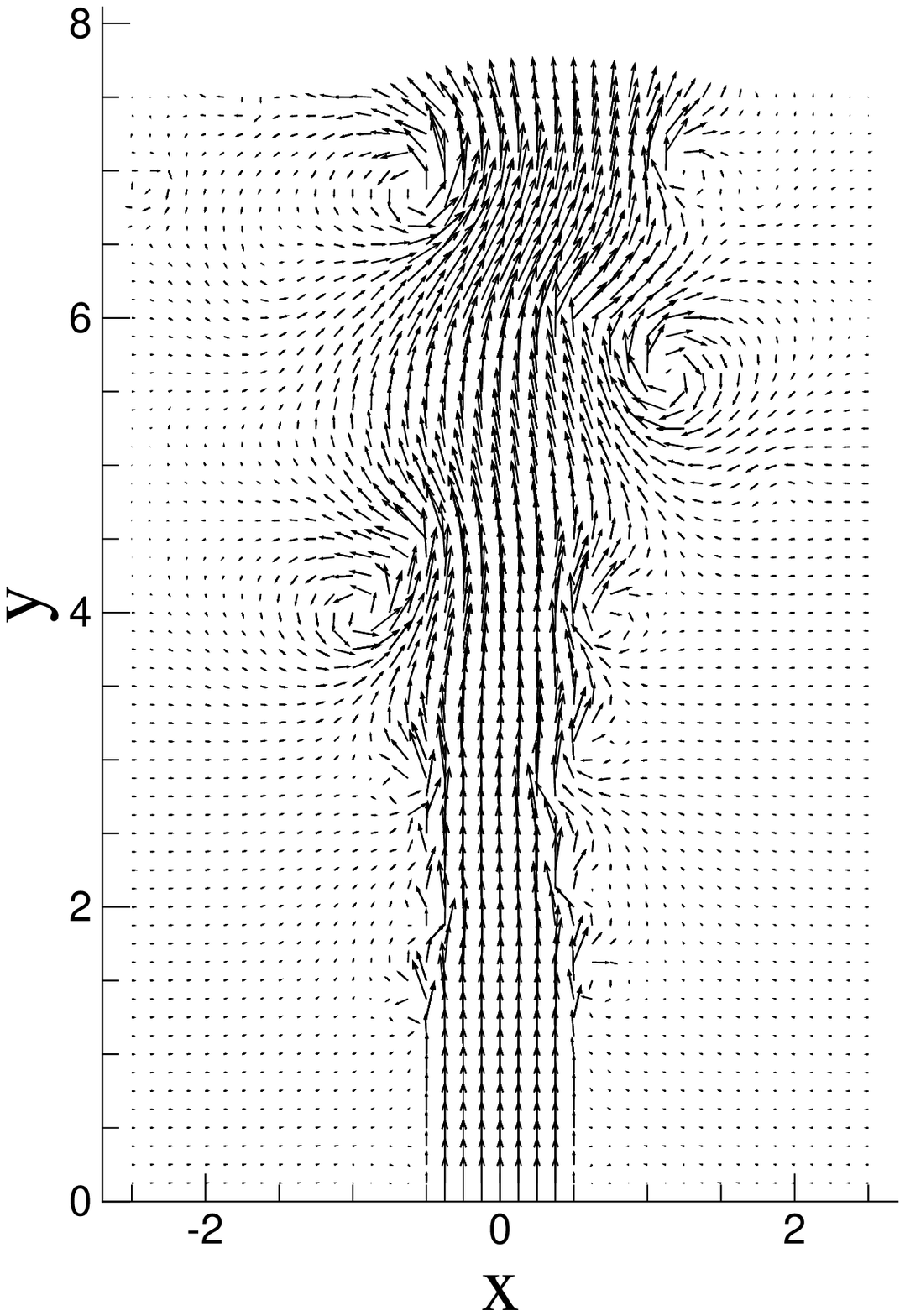}(c)
}
\caption{
Jet in open domain:
snapshots of instantaneous velocity fields at
(a) $Re=2000$,
(b) $Re=5000$, and
(c) $Re=10000$.
Velocity vectors are plotted on every nineth quadrature 
points in each direction within each element.
Results are obtained with $D_0=\frac{1}{U_0}$ in 
the open boundary condition.
}
\label{fig:jet_vel}
\end{figure}

The flow characteristics are illustrated 
by snapshots of the instantaneous velocity
shown in Figure \ref{fig:jet_vel}
at these Reynolds numbers.
These results are obtained using $D_0=\frac{1}{U_0}$
in the open boundary condition \eqref{equ:obc_D_3}.
At $Re=2000$, the jet appears to be stable
within a distance of at least $3d$ downstream 
of the inlet (Figure \ref{fig:jet_vel}(a)).
Beyond this region, the jet becomes unstable and
a pair of vortices forms, which travels
downstream along with the jet and eventually
crosses the upper open boundary and
exits the domain.
The velocity field appears 
symmetric with respect to the jet centerline
at this Reynolds number.
As the Reynolds number increases 
to $Re=5000$, 
the stable region immediately downstream of
the inlet becomes shorter (approximately $2d$,
see Figure \ref{fig:jet_vel}(b)),
and the velocity distribution 
has lost the symmetry with respect to
the jet centerline.
Pairs of vortices can be observed to
form, wrapping around the jet at different
downstream locations.
At $Re=10000$,
The stable region downstream of the inlet becomes
evern shorter (about $d$),
and the vortices forming along the jet
are notably more numerous (Figure \ref{fig:jet_vel}(c)).

\begin{figure}
\centering
\includegraphics[width=4.5in]{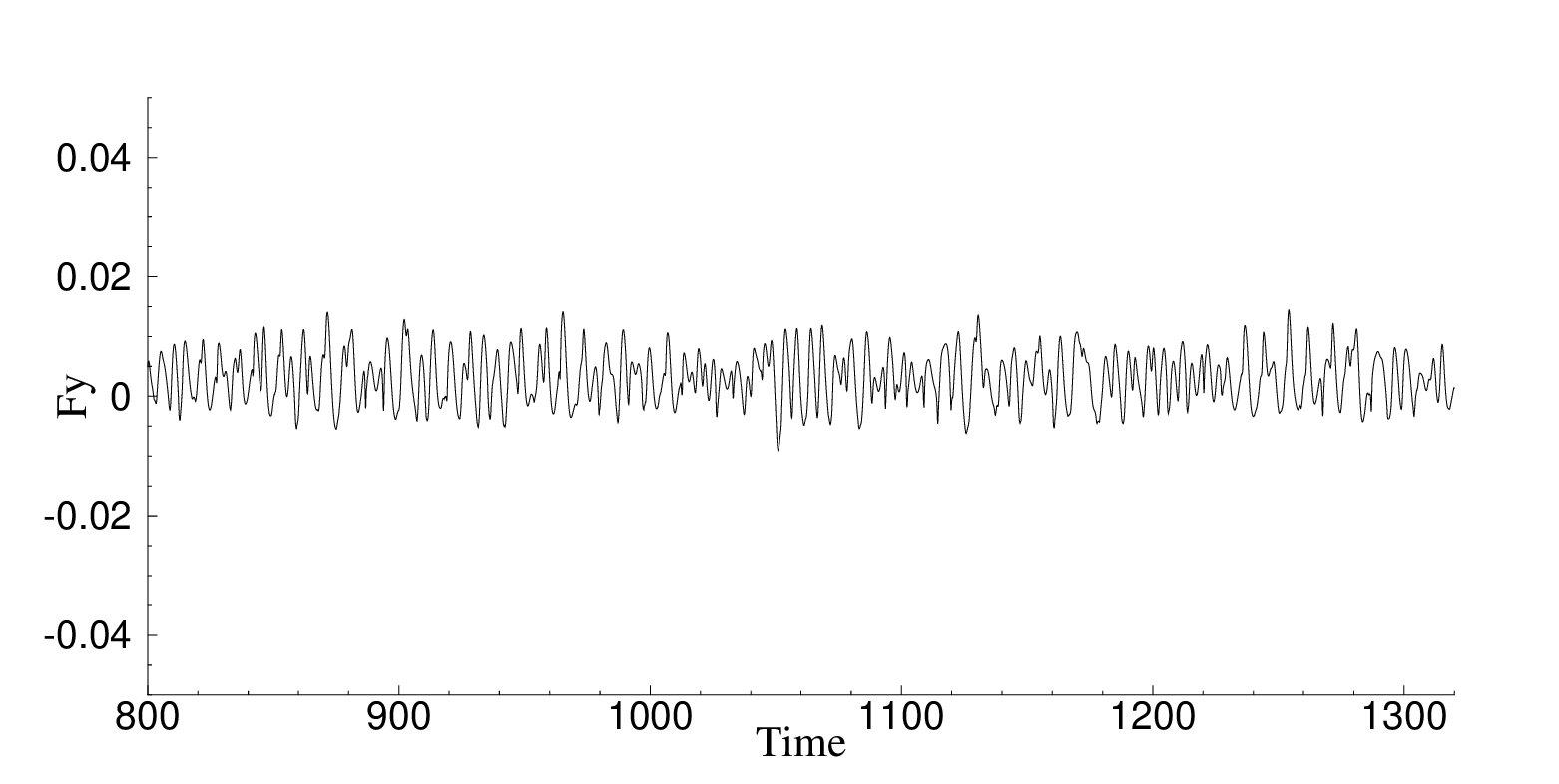}(a)
\includegraphics[width=4.5in]{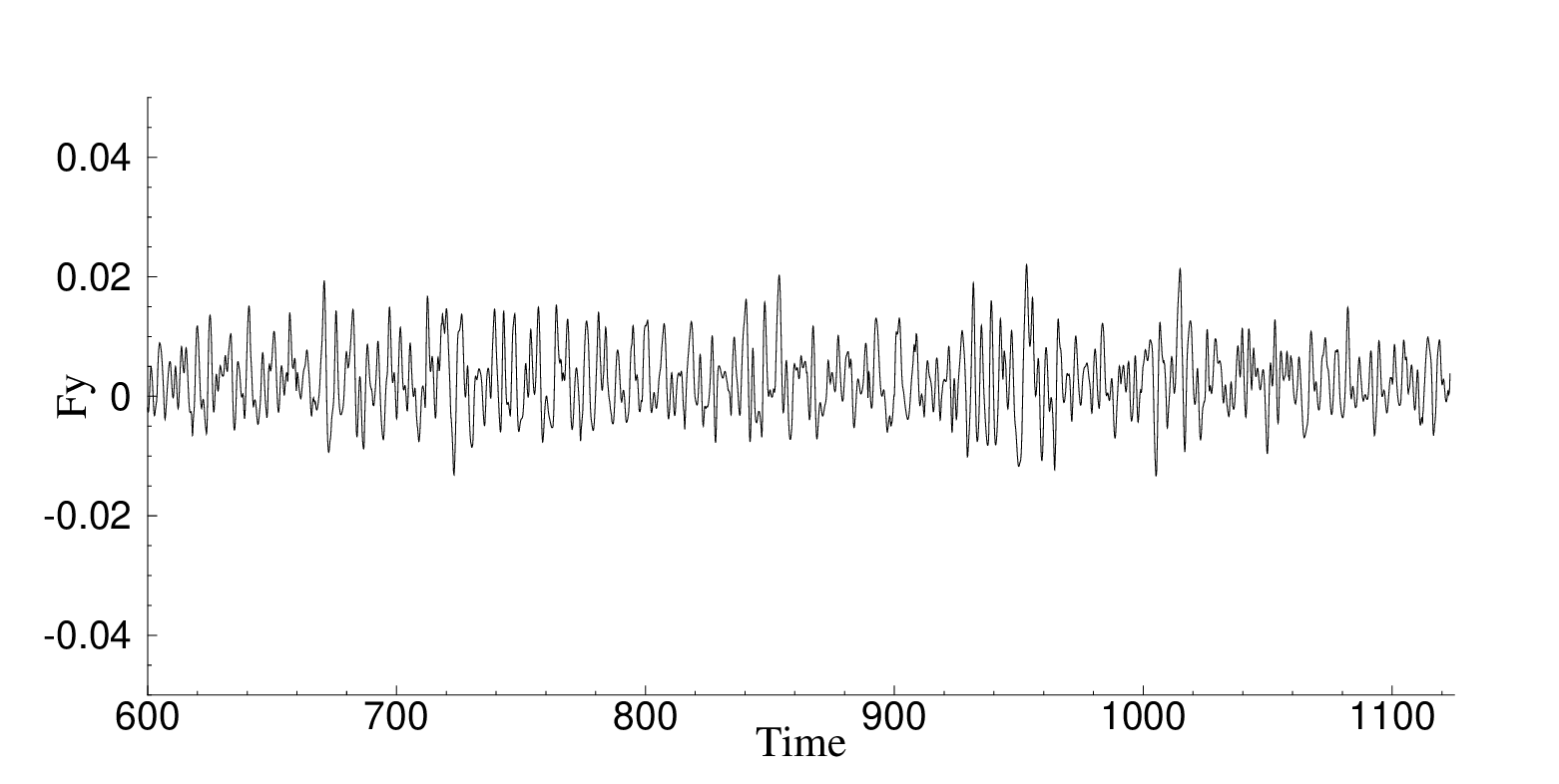}(b)
\includegraphics[width=4.5in]{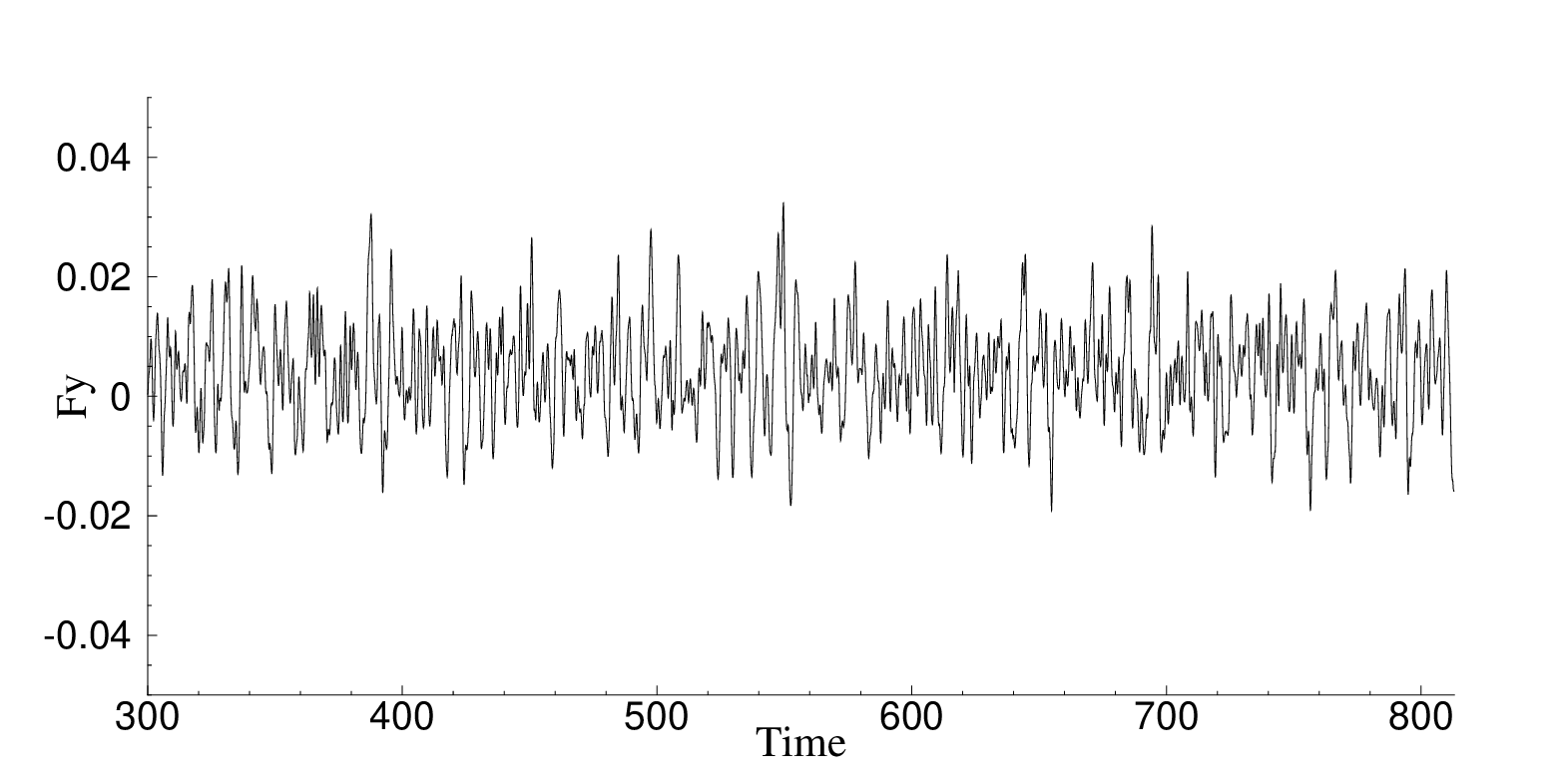}(c)
\caption{
Jet in open domain: Time histories of 
the vertical force component 
exerting on the wall at Reynolds numbers
(a) $Re=2000$,
(b) $Re=5000$, and
(c) $Re=10000$.
Results correspond to $D_0=\frac{1}{U_0}$
in the open boundary condition.
}
\label{fig:jet_fy_hist}
\end{figure}

The jet simulations using the current method
 are long-term stable, even in the presence
of backflows or vortices (see e.g. Figure \ref{fig:jet_vel}(c))
at the open domain boundaries.
This is demonstrated by Figure \ref{fig:jet_fy_hist},
in which we show a window of the time histories
of the vertical force acting on the bottom wall
for these three Reynolds numbers.
Note that the horizontal force ($x$ component)
on the wall is essentially zero.
These results  again correspond to
$D_0=\frac{1}{U_0}$ in the open boundary
condition \eqref{equ:obc_D_3}.
One can observe that the force signals are highly
unsteady, and the fluctuations appear
more energetic and chaotic 
with increasing Reynolds numbers 
(Figure \ref{fig:jet_fy_hist}(c)).
The large temporal window shown here,
over $500\frac{d}{U_0}$ or approximately 
$67$ flow through times,
signifies the long-term stability
of our simulations.

% other OBCs

The term containing $\Theta_0(\mathbf{n},\mathbf{u})$ in the open 
boundary condition \eqref{equ:obc_D_3}
is critical to the stability of the current
simulations. For comparison,
we have also employed the
boundary condition \eqref{equ:convec_like_1}
on the open boundaries for simulations of
the current problem, and observe that
the computation is unstable
for all the Reynolds numbers considered here.
The simulations blow up instantly 
as the vortices reach 
the open boundaries.

%% comparison showing effects of D0 values
%% velocity-pressure distributions
%% tables of forces

\begin{figure}
\centering
\includegraphics[width=4.5in]{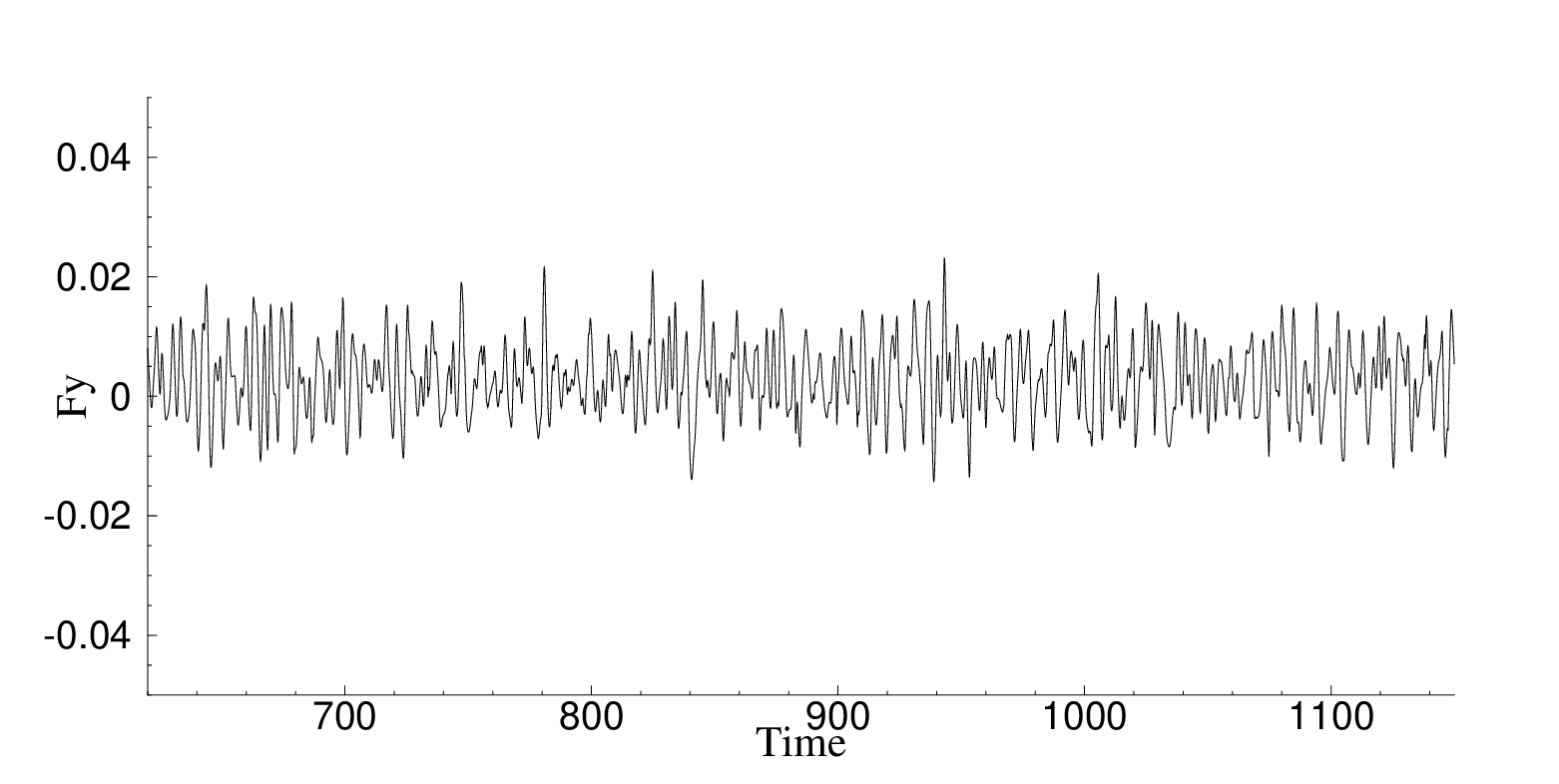}(a)
\includegraphics[width=4.5in]{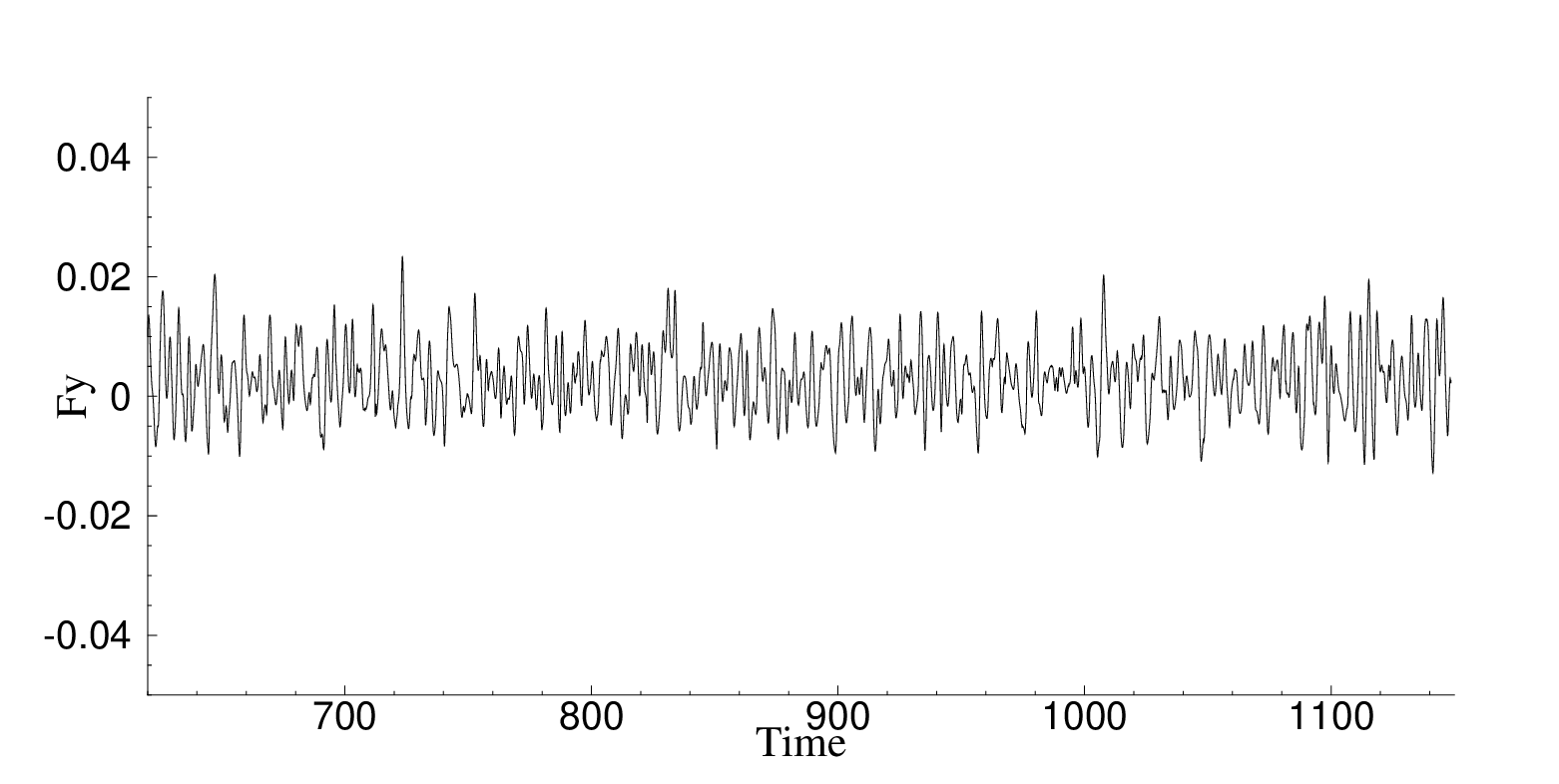}(b)
\caption{
Jet in open domain ($Re=5000$): time histories of the vertical 
force component obtained with different 
$D_0$ values in OBC: (a) $D_0U_0=0$
and (b) $D_0U_0=2.0$. These can be compared with
Figure \ref{fig:jet_fy_hist}(b), which corresponds
to $D_0U_0=1.0$  at the same Reynolds number.
}
\label{fig:jet_fy_hist_D0}
\end{figure}

\begin{table}[t]
\begin{center}
\begin{tabular*}{0.6\textwidth}{ @{\extracolsep{\fill}} l c c c }
\hline
$D_0U_0$ & $\overline{f}_x$ & $\overline{f}_y$ & $f_y^{\prime}$ \\
$0.0$ & $-1.621E-6$ & $2.891 E-3$ & $6.065E-3$ \\
$0.5$ & $1.203E-6$ & $2.864E-3$ & $5.713E-3$ \\
$1.0$ & $-9.599E-7$ & $2.729E-3$ & $5.549E-3$ \\
$2.0$ & $2.829E-7$ & $2.910E-3$ & $5.541E-3$ \\
\hline
\end{tabular*}
\end{center}
\caption{
Mean and rms forces on the wall for the jet 
problem at $Re=5000$ obtained using different
$D_0$ values in OBC. 
$\overline{f}_x$ and $\overline{f}_y$
denote the time-averaged mean forces in 
horizontal and vertical directions, respectively.
$f_y^{\prime}$ is the rms force
in the vertical direction.
}
\label{tab:jet_force}
\end{table}

The results shown so far have been obtained using
$D_0=\frac{1}{U_0}$ in the open 
boundary condition \eqref{equ:obc_D_3}.
We next concentrate on the Reynolds number
$Re=5000$, and look into the effect of $D_0$
 on the simulation results.
We have considered several $D_0$
values in the open boundary condition:
$D_0U_0=0$, $0.5$, $1.0$, and $2.0$.
Figure \ref{fig:jet_fy_hist_D0}
shows time histories of the vertical force
on the wall at $Re=5000$
obtained using $D_0=0$ and $D_0=\frac{1}{U_0/2}$
in the open boundary condition \eqref{equ:obc_D_3}.
These plots can be compared with
Figure \ref{fig:jet_fy_hist}(b),
which corresponds to $D_0=\frac{1}{U_0}$
at the same Reynolds number.
These force signals appear qualitatively similar.

To provide a quantitative comparison,
we have listed in Table \ref{tab:jet_force}
the time-averaged mean and rms
forces acting on the wall corresponding
to different $D_0$ values in the open boundary
condition. 
The mean force in the horizontal direction
($\overline{f}_x$) is essentially zero for all
cases.
Both the mean and rms forces in the vertical
direction obtained using different
$D_0$ values are close,
with a maximum difference of about 
$6.6\%$ for the mean vertical force
and a maximum difference of about $9.5\%$
for the rms vertical force.
The $D_0$ effect on the global flow
quantities (such as forces)
observed here for the jet flow 
is approximately in line
with the observations for the cylinder
flow in Section \ref{sec:cylinder}.
But the differences observed for this problem appear 
slightly larger.

% what else to discuss here?

%% discharge of vortices across outflow boundaries

%% Summary, discuss computational cost briefly
\section{Concluding Remarks}
\label{sec:summary}

% what have I done in this paper?
% what are the results?
% what do the results mean?
% what is the importance?
% what is the relation between this OBC 
%   and traction-free OBC and convective OBC?
%

The main contributions of the current paper
are two-fold:
\begin{itemize}

\item
We have presented a new type of energy-stable
open boundary condition for incompressible flow
simulations. This boundary condition
ensures the energy stability of the system,
and in some sense can be analogized to the usual convective
boundary condition.
The current open boundary condition
can be reduced to that of \cite{DongS2015}
if the inertia term
involved herein vanishes (by setting $D_0=0$).
Note that the current boundary condition 
can be generalized to 
a family of convective-like energy-stable
 open boundary conditions
with an idea similar to that employed
in \cite{DongS2015}; see the remarks in 
Section \ref{sec:obc}.

\item
We have presented an efficient 
numerical algorithm for 
the proposed open boundary condition.
The key issue here lies in the numerical treatment
of the inertia term involved
in the current open boundary condition.
Our algorithm combines a rotational
velocity-correction type splitting strategy
for the Navier-Stokes equations
with two different implicit approximations
of the inertia term
in the open boundary condition
for the pressure and velocity sub-steps.
The algorithm leads to  
Robin-type conditions for both the discrete pressure and
the discrete velocity on the outflow/open boundary.

\end{itemize}

% relation between current OBC and convective OBC
%   and traction-free OBC

We would like to point out that 
the open boundary condition proposed herein
can be considered as the energy-stable
version of a combined traction-free and
convective boundary condition (see equation
\eqref{equ:convec_like_1}).
As discussed in Section \ref{sec:obc},
this condition will be reduced to the traction-free
condition if the inertia
term vanishes (i.e. $D_0=0$),
and if $p=0$ is imposed on
the outflow boundary it will be  
reduced essentially to the 
usual convective boundary
condition.

% what is the advantage of the current OBC?
% (1) smoother flow pattern at O-boundary
% (2) no need for projection to H^1(\partial\Omega) space
%     in implementation
%

The method developed in the current work 
for dealing with outflow/open boundaries
exhibits favorable properties
when compared with those of \cite{DongKC2014,DongS2015}
in at least two aspects.
First, it can lead to qualitatively smoother flow patterns
at or near the outflow boundary, because
the current boundary condition (when $D_0> 0$) provides a
control over 
the velocity on the outflow boundary.
This has been demonstrated by the numerical simulations
of Section \ref{sec:tests}.
Second,
it provides a simpler implementation (with $D_0>0$) with
$C^0$ spectral-element and finite-element type
methods. This is because 
the current method essentially imposes
a discrete Robin-type condition for
both the velocity and the pressure on the
outflow boundary, and requires only 
an update to the coefficient matrix by
a surface integral in the implementation.
In contrast, with the methods of \cite{DongKC2014,DongS2015}
a pressure Dirichlet type condition is imposed
on the outflow boundary, and a projection
to the $H^1(\partial\Omega_o)$ space of
the pressure Dirichlet data will be
required with $C^0$ elements in the implementation.
This projection is more involved than 
the evaluation of the surface integral
required by the current method.

% what are the results?

The effectiveness of the current method 
has been demonstrated by extensive 
numerical simulations.
Comparison with the experimental data
shows the accuracy of the current method.
At higher Reynolds numbers
when backflows and strong vortices
occur at the outflow/open boundaries,
numerical results have demonstrated
the long-term stability 
using the current method. 
It is observed 
that the method allows the vortices 
to discharge from the domain in
a fairly natural fashion, even at
quite high Reynolds numbers (up to $Re=10000$
tested here).

% what are implications?
% what is the importance?
% what do the results mean?

We anticipate that the method developed 
herein will be instrumental in
numerical studies of wakes, jets,
shear layers, and other types of flows involving
physically unbounded domains, 
especially for high Reynolds numbers.
It would be very interesting to 
extend the idea 
and develop analogous boundary conditions for
moving domains such as in an arbitrary Lagrangian
Eulerian context.
This would be important and useful to
applications such as vortex/flow induced
vibrations.
Future research will address such
and related problems.

% what else to discuss here?

%%
\section*{Acknowledgement}
This work is partially supported by 
 NSF (DMS-1318820) and 
ONR (N000141110028).

%%
%\input Appendix

%%% Figures
%\input figures

%\newpage
%
\bibliographystyle{plain}
\bibliography{obc,mypub,nse,sem,cyl}
%contact_line,interface,mypub,basis,nse,time_integration,sem}
%pfem,time_integration,mypub,sem,taylor_couette,precon}

\end{document}